\documentclass{jfm}

\usepackage[]{amsmath}
\usepackage[]{physics}
\usepackage[]{xcolor}
\usepackage[]{subfigure}
\usepackage{upgreek}
\usepackage{overpic}
\usepackage[]{multirow}
\usepackage[]{url}
\usepackage[]{nicefrac}
\usepackage{bm}

\graphicspath{{imgs/}}

\DeclareMathAlphabet{\mathcal}{OMS}{cmsy}{m}{n}
\DeclareMathAlphabet\mathbfcal{OMS}{cmsy}{b}{n}
\DeclareMathOperator*{\minimize}{minimize~}
\DeclareMathOperator*{\subjectto}{subject~to~}

\renewcommand{\ip}[2]{\left( #1, #2 \right)_\Omega }

\begin{document}

\newtheorem{lemma}{Lemma}
\newtheorem{corollary}{Corollary}

\shorttitle{Nonlinear correlations in reduced-order modeling} %for header on odd pages
\shortauthor{J. L. Callaham, and S. L. Brunton, J.-Ch.~Loiseau} %for header on even pages

%\title{Reduced-order modeling and nonlinear correlations : why does it matter?}
\title{On the role of nonlinear correlations in reduced-order modeling}

\author{Jared L. Callaham\aff{1}, Steven L. Brunton\aff{1}, Jean-Christophe Loiseau\aff{2}}

\affiliation{
	\aff{1}
	Department of Mechanical Engineering, University of Washington,
    Seattle, WA 98195, USA
	\aff{2}
	Arts et M\'{e}tiers Institute of Technology, CNAM, DynFluid, HESAM Universit\'{e}, F-75013 Paris, France
}

\maketitle

\begin{abstract}
A major goal for reduced-order models of unsteady fluid flows is to uncover and exploit latent low-dimensional structure.
%Despite significant progress, constructing a minimal description of the dynamics remains challenging.
Proper orthogonal decomposition (POD) provides an energy-optimal linear basis to represent the flow kinematics, but converges slowly for advection-dominated flows and tends to overestimate the number of dynamically relevant variables.
We show that nonlinear correlations in the temporal POD coefficients can be exploited to identify the underlying attractor, characterized by a minimal set of driving modes and a manifold equation for the remaining modes.
By viewing these nonlinear correlations as an invariant manifold reduction, this least-order representation can be used to stabilize POD-Galerkin models or as a state space for data-driven model identification.
In the latter case, we use sparse polynomial regression to learn a compact, interpretable dynamical system model from the time series of the active modal coefficients.
We demonstrate this perspective on a quasiperiodic shear-driven cavity flow and show that the dynamics evolve on a torus generated by two independent Stuart-Landau oscillators.
These results emphasize the importance of nonlinear dimensionality reduction to reveal underlying structure in complex flows.
\end{abstract}

%%%%%%%%%%%%%%%%%%%%%%%%%%%%%%%%
%%%%%     INTRODUCTION     %%%%%
%%%%% %%%%%%%%%%%%%%%%%%%%%%%%%%
\section{Introduction}
\label{sec: introduction}

Many systems with complex, multiscale structure are nevertheless characterized by emergent large-scale coherence~\citep{Haken1983, Cross1993}, generating low-dimensional structure often conceptualized as an attracting or slow manifold.
This phenomenon is especially relevant in fluid dynamics, where successive bifurcations lead to increasingly complex behavior and eventually the transition to turbulence~\citep{Landau1944, Stuart1958jfm, jas:lorenz:1963, Ruelle1971cmp, SwinneyGollub}.
Dynamical models that capture this intrinsic low-dimensional structure can improve our physical understanding and are critical for real-time optimization and control objectives~\citep{book:noack:2011, amr:brunton:2015,Rowley2017}.

\begin{figure}
\centering
\includegraphics[width=\textwidth]{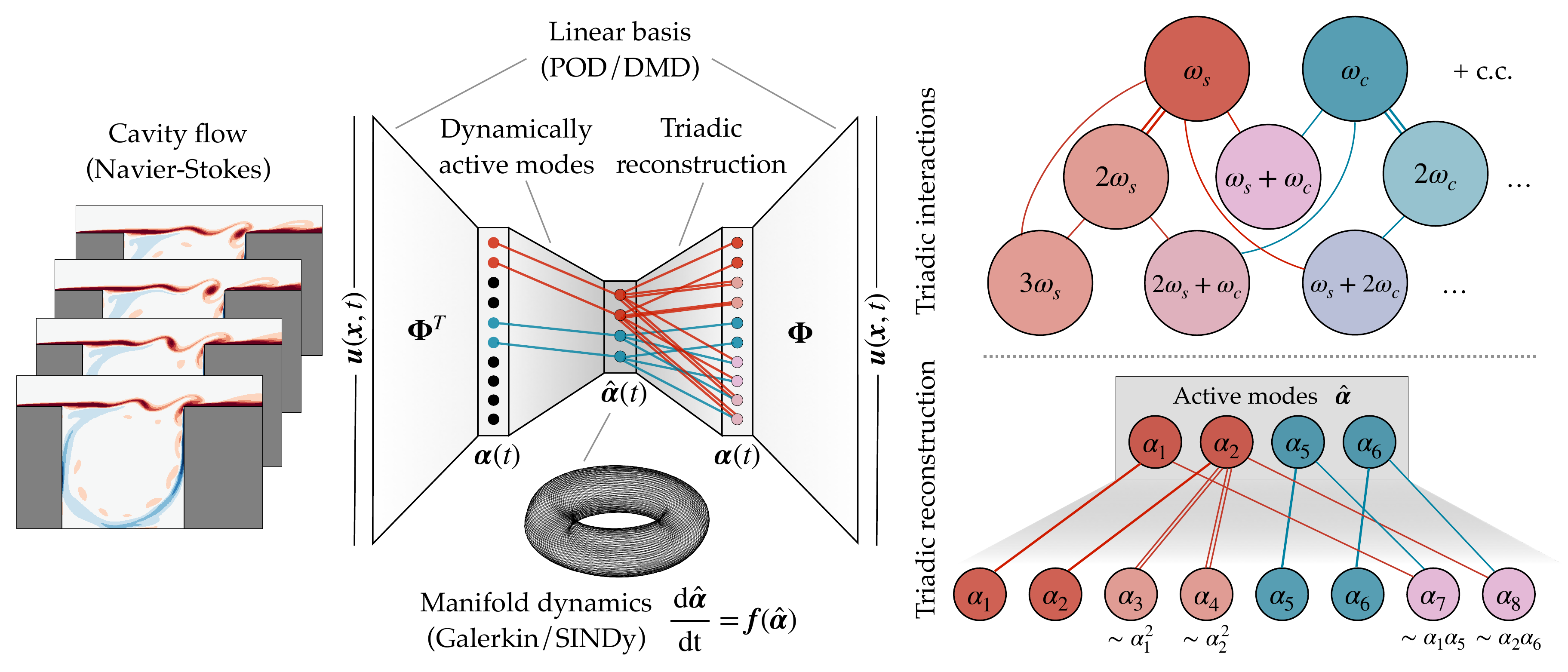}
\caption{
Schematic of the model reduction approach exploiting nonlinear correlations.
The flow fields are first projected onto a linear modal basis $\bm \Phi$, yielding modal coefficients $\bm \alpha(t)$.
The quasiperiodic dynamics can be described by four degrees of freedom; the rest of the modal coefficients can then be reconstructed with polynomial functions consistent with triadic interactions in the frequency domain.
The dynamics of the active degrees of freedom can be modeled either by restricting the POD-Galerkin dynamics to the toroidal manifold or by identifying a simple, interpretable dynamical system with the SINDy algorithm.
}
\label{fig: intro -- overview}
\end{figure}

Close to a bifurcation, the dynamics are approximately restricted to the manifold described by the amplitudes of the unstable eigenmodes.
The evolution equations for these effective coordinates are given by the normal form for the bifurcation~\citep{GuckenheimerHolmes}, the form of which can be deduced with symmetry arguments~\citep{Golubitsky1988, Glaz2017jfm, Deng2020jfm}, weakly nonlinear analysis~\citep{Stuart1958jfm, jfm:sipp:2007, Meliga2009jfm}, or a center manifold reduction~\citep{jfm:carini:2015}.
Normal forms can describe a wide range of stereotypical dynamics, including bistability, self-sustained oscillations, and chaos.

These arguments are only valid near the bifurcation, although empirical methods can generalize this approach beyond the point where models can be derived via asymptotic expansions.
These methods typically represent the field as a linear combination of modes~\citep{aiaa:taira:2017}, followed by either Galerkin projection onto the governing equations~\citep{jfm:aubry:1988, book:holmes:1996, jfm:noack:2003, book:noack:2011}, data-driven system identification~\citep{pnas:brunton:2016, jfm:loiseau:2018a, Loiseau2019tcfd, Rubini2020jfm}, or a hybrid of the two~\citep{Mohebujjaman2018ijnmf, Xie2018siam}.

A linear modal basis is typically derived as the solution to some optimization problem.
For example, proper orthogonal decomposition (POD) modes minimize the kinetic energy of the unresolved fluctuations for a given basis size, with the residual monotonically decreasing with the basis dimensionality~\citep{book:holmes:1996}.
On the other hand, dynamic mode decomposition (DMD) incorporates temporal information via the spectral decomposition of a best-fit linear evolution operator that advances the flow measurements forward in time~\citep{jfm:schmid:2010,Rowley2009,jcd:tu:2014,book:kutz:2016}.
DMD can also be viewed as a special case of a Koopman mode decomposition, which is based on a spectral analysis of the evolution operator for nonlinear observables~\citep{jfm:rowley:2009,arfm:mezic:2013, Brunton2021koopman}.
Regardless of the optimization problem, linear representations of convection-dominated flows fundamentally suffer from a large Kolmogorov $n$-width, or a linear subspace that slowly approaches full kinematic resolution with increasing dimension~\citep{Grimberg2020jcp}.
In this case, even when enough modes are retained to reconstruct the flow field, the Galerkin model may not faithfully represent the underlying physics.

Reduced-order models based on heavily truncated linear representations are therefore known to suffer from severe instabilities without careful closure modeling~\citep{jfm:aubry:1988, Noack2008jnet, cmame:wang:2012, Maulik2019jfm}.
From a numerical perspective, part of the problem is the Galerkin formulation commonly used to derive a set of time-continuous ordinary differential equations~\citep{jcp:carlberg:2017, Grimberg2020jcp}, but there are also at least two physical reasons for the instability of projection-based models:

\begin{enumerate}
    \item The higher-order modes tend to represent smaller scales of the flow, which are responsible for the bulk of energy dissipation, so that truncated models may not accurately capture the energy cascade in the flow.
    This motivates eddy viscosity-type modifications by analogy with classical turbulence closure models~\citep{jfm:aubry:1988, cmame:wang:2012} as well as alternative Galerkin schemes that explicitly target energy balance~\citep{jfm:balajewicz:2013, Mohebujjaman2017jcp}.
    \item The dimensionality of the linear subspace required to reconstruct the flow field may significantly exceed the intrinsic dimension of the attractor of the system.
    Since traditional model reduction methods have one state variable per mode, the projected dynamics may have many more degrees of freedom than the physical system.
    For instance, the traveling wave solution to a linear advection equation on a periodic domain may require arbitrarily many Fourier modes for a linear reconstruction, yet with the method of characteristics the state is determined by a single degree of freedom: the scalar phase.
\end{enumerate}

The competition between these two effects tends to lead to fragile Galerkin systems without further modeling.
Enough modes must be retained to sufficiently resolve dissipation, but this large number of kinematic modes may be considerably larger than the number of dynamic degrees of freedom.  
Therefore, the dynamics of models that include a large number of modes may not resemble those of the underlying flow.
A case study of these considerations is the pioneering work of~\citet{jfm:noack:2003} modeling the two-dimensional flow past a cylinder.
With an augmented POD basis and a careful dynamical systems analysis, they reduce a structurally unstable eight-dimensional Galerkin system to a two-dimensional cubic model that reproduces the dominant flow physics.

%% Nonlinear dimensionality reduction
The issues of stability and validity are intimately connected to the question of correlation.
The temporal coefficients of POD modes are linearly uncorrelated on average~\citep{book:holmes:1996}, but no such guarantee is available for nonlinear correlation.
For example, one mode may be a harmonic of another; in this case their temporal coefficients are linearly uncorrelated but the harmonic is a perfect algebraic function of the fundamental.
If these coefficients are modeled independently, as in a classical Galerkin system, slight inaccuracies can lead to decoherence and unphysical solutions.
%Since DMD modes by nature approximate oscillatory behavior, these effects can be even more pronounced when the field is represented in a DMD basis.

In this work we show that nonlinear correlations can be exploited to construct stable and accurate low-dimensional models without closure assumptions, as shown schematically in figure~\ref{fig: intro -- overview}.
After projecting data from a direct numerical solution of a quasiperiodic shear-driven cavity onto a basis of DMD modes, the recently proposed randomized dependence coefficient~\citep{rdc:lopez-paz:2013} allows us to clearly distinguish the active degrees of freedom from correlated higher harmonics and nonlinear crosstalk.
In this minimal representation, the dynamics occur on a 2-torus, while the rest of the modes, which arise as triadic interactions of the active variables in the frequency domain, can be expressed as polynomial functions of the dynamically active variables.
The restriction to this manifold stabilizes a standard POD-Galerkin model, avoiding both decoherence and energy imbalance.
This representation is also a natural basis for data-driven system identification methods; we apply the sparse identification of nonlinear dynamics (SINDy) algorithm~\citep{pnas:brunton:2016} and show that the flow can be accurately described by two independent Stuart-Landau equations.

This work is organized as follows.
In Section~\ref{sec: example} we use two model PDEs to give brief analogies motivating our use of nonlinear correlation.
We introduce the open cavity flow and direct numerical simulation in Section~\ref{sec: flow configuration} and give POD and DMD analyses in Section~\ref{sec: modes}.
Section~\ref{sec: rom} introduces the reduced-order modeling techniques of Galerkin projection and SINDy.
In Section~\ref{sec: correlations} we show how nonlinear correlations arise in the modal analysis of the flow and how this can be exploited for the reduced-order models.
A comparison and analysis of the various models is given in Section~\ref{sec: results}, followed by a final discussion in Section~\ref{sec: discussion}.

%%%%%%%%%%%%%%%%%%%%%%%%%%%%%%%
%%%%%     TOY PROBLEM     %%%%%
%%%%%%%%%%%%%%%%%%%%%%%%%%%%%%%
%\section{Galerkin projection for a simple hyperbolic problem}
%\section{The linear dispersion relation as nonlinear correlation}
\section{The origins of nonlinear correlation}
\label{sec: example}

Many features of projection-based models of advection-dominated flows are demonstrated by simple scalar PDEs.
In particular, limitations of the Galerkin representation of hyperbolic problems can be seen in the linear constant-coefficient advection equation, while Burgers' equation is a minimal example of the key role of nonlinearity in the full Navier-Stokes equations.

\subsection{The linear dispersion relation as nonlinear correlation}
\label{sec: example -- advection}

One of the fundamental reasons that Galerkin models of advection-dominated flows tend to be fragile is that they introduce additional variables that do not correspond to physical degrees of freedom.
This is perhaps illustrated most clearly by the linear advection equation on a periodic domain:
\begin{equation}
    u_t + c u_x = 0, \hspace{1cm} x \in (0, L).
\end{equation}
For any initial condition $u(x, 0) = u_0(x)$, this equation has the simple traveling wave solution $u(x, t) = u_0(x - ct)$.
Given the initial condition, the only effective degree of freedom is the phase $ct/L$.
However, the problem could also be solved by means of a Fourier expansion
\begin{equation}
\label{eq: example -- fourier}
    u(x, t) = \sum_{n=-\infty}^{\infty} a_n(t) e^{ik_nx}
\end{equation}
with $k_n = 2\pi n/L$
The Galerkin system in this orthogonal basis (see Sec.~\ref{sec: rom}) is
\begin{equation}
\label{eq: example -- adv-galerkin}
    \dot{a}_n(t) = -i\omega_n a_n(t), \hspace{1cm} \omega_n = k_n c n = 0, \pm 1, \pm 2, \dots
\end{equation}
The relationship between frequency $\omega$ and wavenumber $k_n$ is the \textit{dispersion relation}; in this case it implies that all scales are carried at the same speed $c$.

\begin{figure}
     \centering
     \includegraphics[width=1\textwidth]{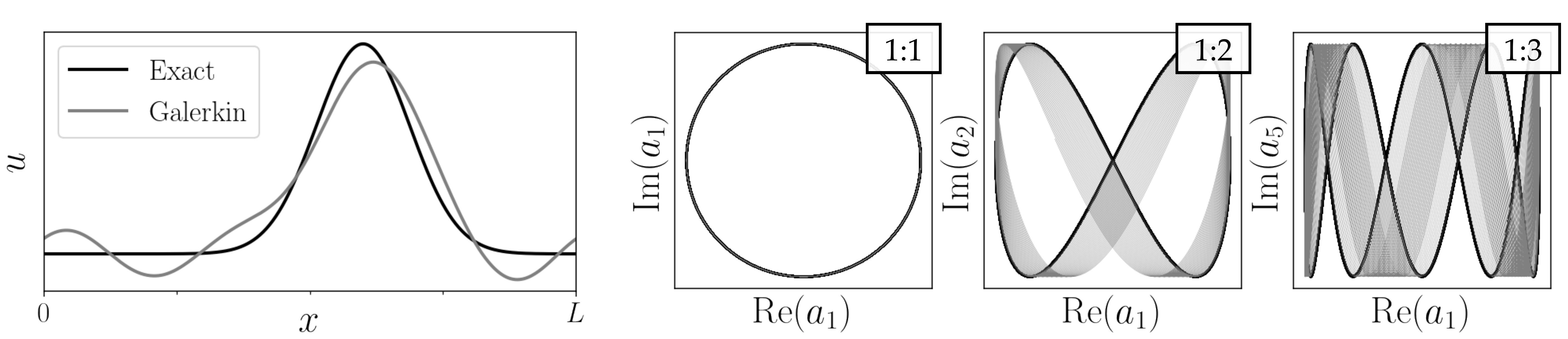}
     \caption{Linear advection equation with errors $\epsilon_n \sim \mathcal{N}(0, \epsilon^2)$ in the dispersion relation $\omega_n = c k_n$.
     The Galerkin model (grey) loses coherence with the exact solution (black) over a time scale $1/\epsilon$.
     If the polynomial correlations implied by the dispersion relation are enforced explicitly, the model is robust to such errors.
     Nonlinear correlation in the true system, given by~\eqref{eq: example -- nonlinear-correlation}, appears in the Lissajous-type phase portraits of the Fourier coefficients (right).
     Similar behavior manifests in Galerkin models of nonlinear advection-dominated flows.}
     \label{fig: advection}
   \end{figure}

With this analytic dispersion relation,~\eqref{eq: example -- adv-galerkin} is equivalent to the traveling wave solution, since $a_n(t) = e^{-i\omega_n t} a_n(0)$:
\begin{equation}
    u(x, t) = \sum_{n=-\infty}^{\infty} a_n(0) e^{ik_n(x-ct)} = u_0(x - ct).
\end{equation}
However, the Galerkin model has introduced many degrees of freedom in the harmonics $a_n$ by artificially separating space and time.
If the projection is approximated numerically or with empirical basis modes, the estimated system may include some error, so that $\omega_n = k_n c + \epsilon_n$.
In this case the Galerkin system will be dispersive, i.e. each wavenumber will propagate with a slightly different speed.
The traveling wave solution will tend to lose coherence on a timescale $1/\epsilon$, as shown in Figure~\ref{fig: advection}.

An alternative perspective on the dispersion relation is that it specifies nonlinear correlations between the temporal coefficients $a_n$, removing the spurious degrees of freedom introduced by Galerkin projection.
The linear dispersion relation $\omega_n = n k_1 c$ implies the nonlinear relationship for harmonics
\begin{equation}
\label{eq: example -- nonlinear-correlation}
    a_n(t) = e^{-in k_1 c t} a_n(0) \propto a_1^n,
\end{equation}
with the proportionality determined by the initial condition.
Then the only degree of freedom is $a_1$, and the traveling wave solution is recovered by the Galerkin model projected onto this mode.
In dynamical systems terminology, the solution is restricted to a one-dimensional manifold: a circle representing the phase of the leading Fourier coefficient.
In this case the decoherence does not lead to instability because the system is purely linear with purely imaginary eigenvalues, but in nonlinear systems with nonzero linear growth rates the departure from the solution manifold can be catastrophic.

\subsection{Triadic interactions and the energy cascade}
\label{sec: example -- burgers}
   
For more general linear systems the preceding analysis is complicated by non-normality and physical dispersion, and the concept of a dispersion relation is not well-defined for nonlinear dynamical systems.
Nevertheless, analogous concepts are similarly important in models of nonlinear PDEs.
For example, Burgers' model is a paradigmatic scalar conservation equation illustrating many features of gas dynamics and nonlinear flows more broadly.
Burgers' equation with viscosity $\epsilon$ is
\begin{equation}
\label{eq: example -- burgers}
    \pdv{u}{t} +  u \pdv{u}{x} = \epsilon \pdv[2]{u}{x}, \hspace{1.5cm} x \in (0, 2\pi).
\end{equation}
On a periodic domain, we can apply the same Fourier expansion~\eqref{eq: example -- fourier} with $L=2\pi$, leading to the Galerkin ODE system
\begin{equation}
\label{eq: example -- burgers-galerkin}
    \dot{a}_k = -\epsilon k^2 a_k -i k \sum_{\ell=-\infty}^\infty a_\ell a_{k-\ell}, \hspace{1cm} k = 0, \pm 1, \pm 2, \dots
\end{equation}

The two right-hand side terms in~\eqref{eq: example -- burgers-galerkin}, originating from the viscous and nonlinear PDE terms respectively, capture several key features of the full Navier-Stokes equations.
First, the convolution-type sum over wavenumbers $\ell$ includes only pairs that sum to $k$; these are the so-called ``triadic'' scale interactions.
Second, it can be shown that the nonlinear term is energy-preserving in the sense that when the energy $a_k a_k^*$ is summed over all wavenumbers the nonlinear term does not contribute to a net change in energy of the system\footnote{A similar result holds for inhomogeneous flows~\citep{jfm:schlegel:2015}.}.
This suggests that the only role of nonlinearity is to transfer energy between scales.
Meanwhile, the dissipation rate of each mode scales quadratically with wavenumber so that the bulk of dissipation occurs at the smallest scales.

The overall picture of the dynamics in the spectral domain is therefore that the nonlinear term transfers energy from the more energetic large scales to the dissipative small scales.
Since~\eqref{eq: example -- burgers-galerkin} is very similar to the spectral form of the momentum equations for isotropic turbulence~\citep{Tennekes1972book}, this ``energy cascade'' is an important feature of real viscous flows as well.
The energy cascade points to another often-discussed issue with Galerkin models: if the system is truncated at a wavenumber $r$ which is not sufficiently large to capture the net dissipation rate, the energy cascade is interrupted and the system of ODEs will overestimate the energy, potentially even becoming unstable.

This issue is fundamentally different from the decoherence discussed in the context of the linear advection equation.
For example, the issue of fine-scale dissipation is also present in the heat equation, given by equations~\eqref{eq: example -- burgers}-\eqref{eq: example -- burgers-galerkin} without the convective nonlinearity.
Whereas the Fourier-Galerkin representation of advection introduces spurious degrees of freedom, this discussion suggests that in the representation of the heat equation \emph{all} coefficients are dynamically important (self-similarity notwithstanding).
The Galerkin system is therefore an ideal representation of the parabolic dynamics of the heat equation, where the fundamental assumption of separation of variables is valid.

The inability of Fourier decomposition and POD to produce efficient representations of traveling wave physics has long been recognized. 
Fundamentally, these decompositions rely on a space-time separation of variables, which is not a valid assumption for traveling waves.  
Many extensions to POD have been developed for translationally invariant systems and systems with other symmetries~\citep{Rowley2000physd,reiss2018shifted,Rim2018juq,Mendible2020tcfd}. 

For general viscous, nonlinear, advection-driven fluid flows, we might expect advection, triadic interactions, and small-scale dissipation to all be relevant as a result of the joint hyperbolic-parabolic structure of the Navier-Stokes equations.
The intrinsic dimensionality of the system and, conversely, the inaptitude of the Galerkin model, may not be \textit{a priori} clear as a result of a complex interplay between these mechanisms.

For example, if the leading degree of freedom $a_1$ tends to oscillate at a frequency $\omega_0$, representing either a standing or traveling wave, then the $a_2$ dynamics include a term of the form $a_1^2 \sim e^{2i\omega_0 t}$.
Similarly, $\dot{a}_3 \propto a_1 a_2 \sim e^{3i\omega_0 t}$.
In the energy cascade picture, these higher-order modes act as forced, damped oscillators and will tend to respond at the forcing frequencies.
In this manner triadic interactions in the wavenumber domain can also give rise to nonlinear correlations in time and triadic structure in the frequency domain.
This effect has recently been exploited for system identification~\citep{Rubini2020jfm} and modal analysis~\citep{Schmidt2020bispectral}.

In analogy with the dispersion relation, these processes may result in latent structure not immediately obvious in the Galerkin representation.
If this structure is ignored, the behavior of the model may depart significantly from that of the underlying system.
For example,~\citet{Majda2000pnas} showed that a truncated Galerkin model of the inviscid Burgers equation tends towards equipartition of energy rather than a physical solution as a result of a catastrophic decoherence mechanism.
Consequentially, in the following sections we argue that nonlinear correlation and manifold restriction plays an important role in the stability and accuracy of reduced-order models of advection-dominated fluid flows.

%%%%%%%%%%%%%%%%%%%%%%%%%%%%%%%%%%%%%
%%%%%     PLANT DESCRIPTION     %%%%%
%%%%%%%%%%%%%%%%%%%%%%%%%%%%%%%%%%%%%
\section{Flow configuration}
\label{sec: flow configuration}

\begin{figure}
   \vspace{.1in}
     \centering
     \includegraphics[width=.7\textwidth]{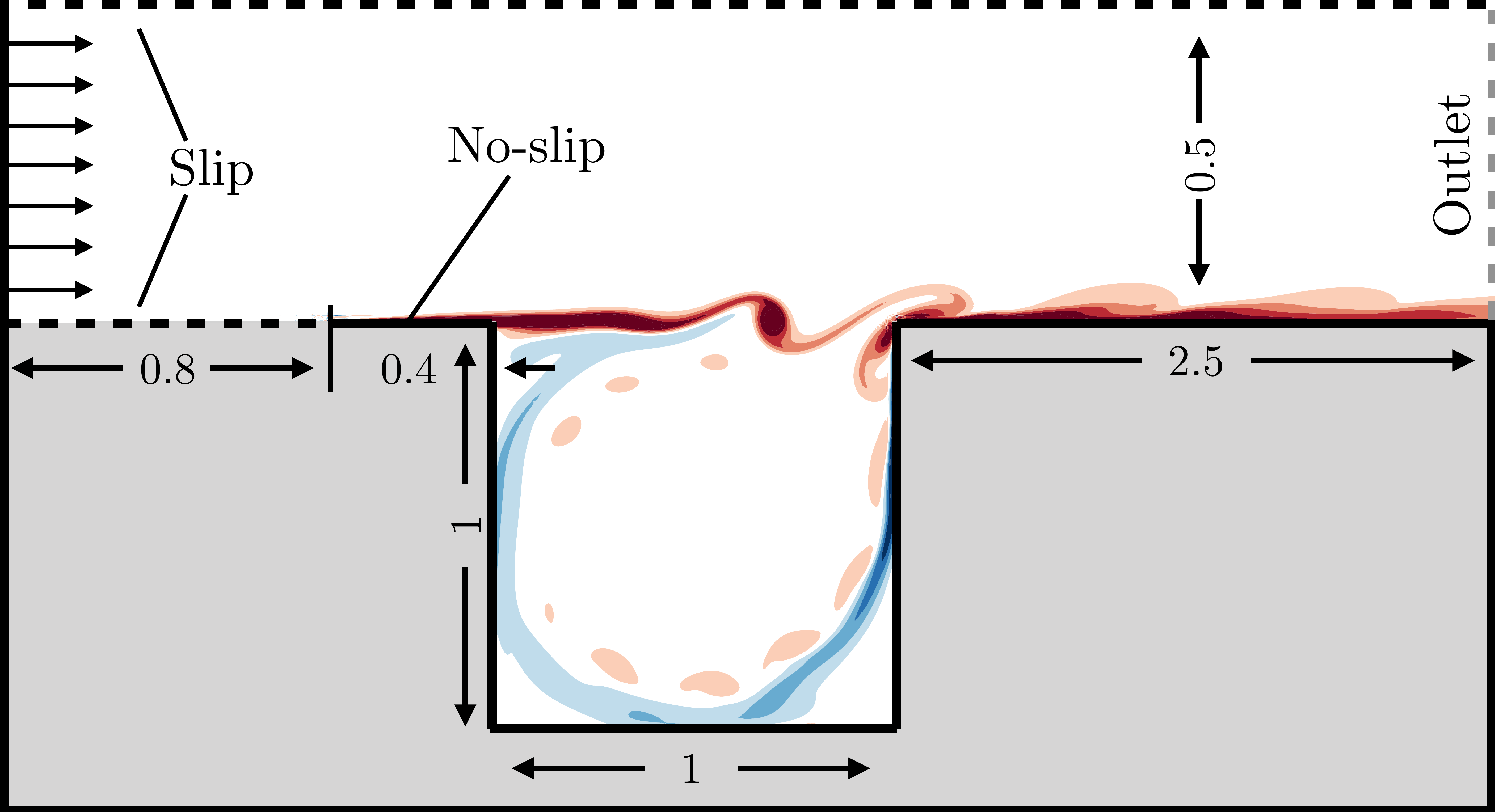}
     \caption{Computational domain and representative instantaneous vorticity field for the shear-driven cavity flow at $Re=7500$ highlighting the vortical structures developing along the shear layer.}
     \label{fig: plant description -- instantaneous visualization}
   \end{figure}

\begin{figure}
    \vspace{.1in}
     \centering
     \includegraphics[width=0.975\textwidth]{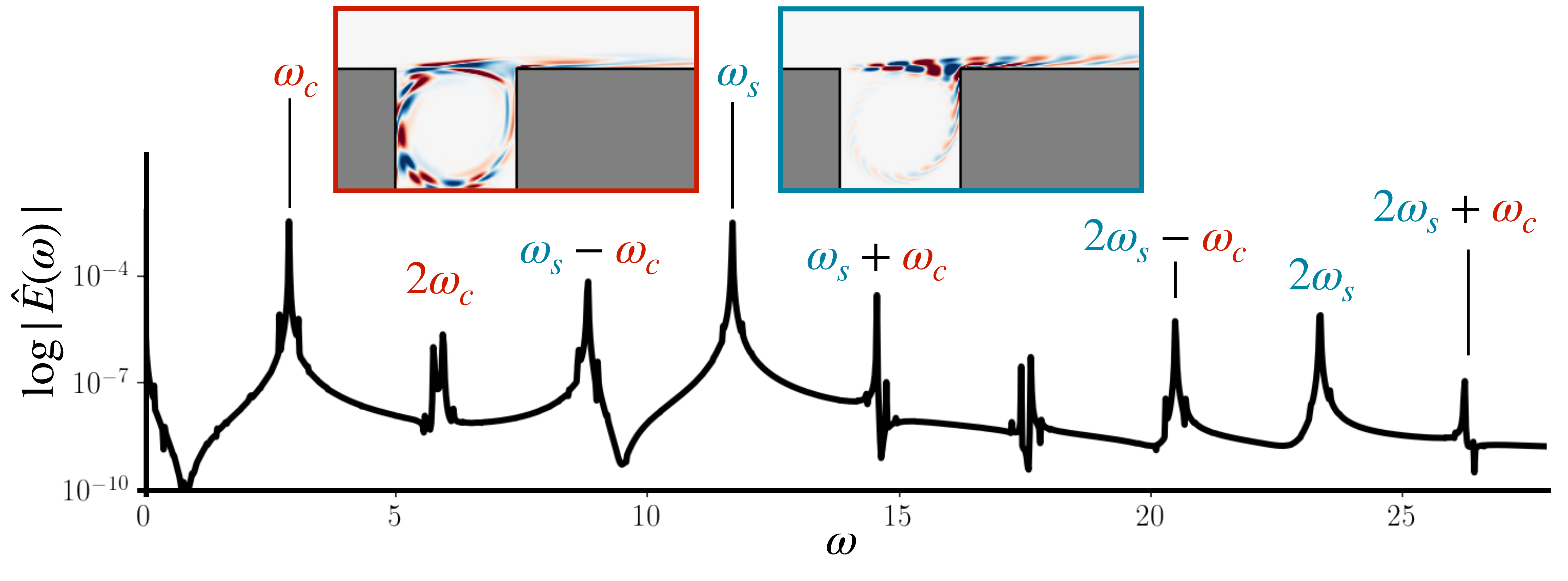}
     \caption{Fourier spectrum $\vert \hat{E} ( \omega ) \vert$ of the fluctuation's kinetic energy at $Re=7500$.
     %The structure of multiple discrete peaks is characteristic of quasiperiodic dynamics (c.f.~\citet{jfm:leclerc:2019}).
     The high-frequency peak ($\omega_s \simeq 12)$ corresponds to the shear layer instability while the low frequency peak ($\omega_c \simeq 3$) are associated to the inner-cavity dynamics. A few other peaks have been labelled based on the quadratic interactions on the two fundamental frequencies for the sake of illustration.
     Multiple closely-spaced peaks are associated with nearby freqeuncy combinations (e.g. $2 \omega_c \approx \omega_s - 2 \omega_c \approx 6$).
     Also shown are the real parts of the DMD modes at $\omega_s$ and $\omega_c$.}
     \label{fig: plant description -- fourier spectrum}
   \end{figure}

  The flow considered in the present work is the incompressible shear-driven cavity flow visualized in Figure~\ref{fig: plant description -- instantaneous visualization}.
  It is a geometrically-induced separated boundary layer flow having a number of applications in aeronautics \citep{joa:yu:1977} or for mixing purposes \citep{jfm:chien:1986}.
  The leading two-dimensional instability of the flow is localized along the shear layer delimiting the outer boundary layer flow and the inner cavity flow \citep{jfm:sipp:2007, amr:sipp:2010}.
  This oscillatory instability relies essentially on two mechanisms.
  First, the convectively unstable nature of the shear layer causes perturbations to grow as they travel downstream.
  %Once they impinge the downstream corner of the cavity, the inner-cavity recirculating flow and the instantaneous pressure feedback then provide the mechanisms allowing these same perturbations to eventually re-excite the upstream portion of the shear layer.
  Once the perturbations impinge upon the downstream corner of the cavity, instantaneous pressure feedback re-excites the upstream portion of the shear layer.
  Coupling of these mechanisms gives rise to a linearly unstable feedback loop at sufficiently high Reynolds numbers ($\Rey_c \gtrsim 4120$, see~\citet{jfm:sipp:2007}).
  A similar unstable loop exist for compressible shear-driven cavity flows, wherein the instantaneous pressure feedback is replaced by upstream-propagating acoustic waves \citep{techreport:rossiter:1964, jfm:rowley:2002, jfm:yamouni:2013}.
  At higher Reynolds numbers, the slowly recirculating flow inside the cavity can also perturb the shear layer.
  This inner cavity mode is similar in spatial structure and oscillation frequency to those observed in two-dimensional lid-driven cavity flows~\citep{Arbabi2017prf}.
  Since the shear layer instability and inner cavity recirculation occur at incommensurate frequencies, the nonlinear coupling between these modes leads to quasiperiodic dynamics, as illustrated in Figure~\ref{fig: plant description -- fourier spectrum}.
  
  Despite its apparent simplicity, this strictly two-dimensional linearly unstable flow configuration has served multiple purposes over the past decade: illustration of optimal control and reduced-order modeling \citep{jfm:barbagallo:2009, jfm:loiseau:2018a, jfm:leclerc:2019}, investigation of the nonlinear saturation process of flow oscillators \citep{jfm:sipp:2007, jfm:meliga:2017}, or as an introduction to dynamic mode decomposition \citep{jfm:schmid:2010}.
  Recent work has also explored the linear stability of its three-dimensional counterpart, in particular the influence of spanwise end-walls~\citep{jfm:liu:2016, jfm:picella:2018}.

  The dynamics of the flow are governed by the incompressible Navier-Stokes equations
  \begin{equation}
    \begin{aligned}
      \displaystyle \frac{\partial \bm{u}}{\partial t} + \nabla \cdot \left( \bm{u} \otimes \bm{u} \right) & = - \nabla p + \frac{1}{Re} \nabla^2 \bm{u} \\
      \nabla \cdot \bm{u} & = 0,
    \end{aligned}
    \label{eq: navier-stokes}
  \end{equation}
  where $\bm{u}(\bm{x}, t) = (u, v)^T$ is the two-dimensional velocity field and $p$ is the pressure field.
  The Reynolds number is set to $\Rey=7500$ based on the free-stream velocity $U_{\infty}$ and the depth $L$ of the open cavity.
  The computational domain and boundary conditions considered herein are the same as in \cite{jfm:sipp:2007, amr:sipp:2010, jfm:loiseau:2018a, Bengana2019jfm}; and \cite{jfm:leclerc:2019}, shown schematically in Figure~\ref{fig: plant description -- instantaneous visualization}.
  
  We perform direct numerical simulaton (DNS) of the flow with the Nek5000 spectral element solver~\citep{nek5000_site}.
  The mesh consists of 6100 eighth-order spectral elements, equivalent to roughly $3.8 \times 10^5$ grid points, refined towards the walls and shear layer.
  The domain is therefore somewhat over-resolved compared to similar studies in order to minimize any numerical errors in the Galerkin projection for higher-order modes.
  Diffusive terms are integrated with third order backwards differentiation, while convective terms are advanced with a third order extrapolation.
  We retain 30000 snapshots from the DNS at sampling rate $\Delta t = 10^{-2}$, a frequency roughly fifty times larger than the high-frequency oscillation of the shear layer.

  Figure \ref{fig: plant description -- instantaneous visualization} depicts an instantaneous vorticity field obtained from direct numerical simulation once the flow has reached a statistical steady state.
  It shows the advection of a vortical structure along the shear layer before it impinges the downstream corner of the cavity.
  One may try to explain such dynamics by means of a linear stability analysis of the steady base flow.
  Unfortunately, since the Reynolds number ($\Rey=7500$) is significantly larger than the critical Reynolds number for the onset of instability, such an analysis only provides limited insights; see Appendix \ref{appendix: linear stability} and~\citet{amr:sipp:2010} for more details.
  %Hence, it appears more convenient to decompose the instantaneous velocity field as the sum of the time-averaged mean flow $\bar{\bm{u}}(\bm{x})$ and a zero-mean fluctuation $\bm{u}^{\prime}(\bm{x}, t)$.
  
  In this case the typical amplitude of fluctuations is not infinitesimal and the associated Reynolds stresses are not negligible.
  The unstable base flow is therefore of limited relevance in the statistically stationary regime and it is more natural to decompose the instantaneous velocity field into a time-averaged mean flow $\bar{\bm{u}}(\bm{x})$ and zero-mean fluctuations $\bm{u}^{\prime}(\bm{x}, t)$.
  For a detailed analysis of the choice between base- and mean-flow expansions, see~\citet{jfm:sipp:2007}.
  
  Figure \ref{fig: plant description -- fourier spectrum} shows the Fourier spectrum of the kinetic energy of the fluctuating component
   \begin{equation}
     E(t) = \displaystyle \frac{1}{2} \int_{\Omega} \bm{u}^{\prime}(\bm{x}, t) \cdot \bm{u}^{\prime}(\bm{x}, t) \ \mathrm{d}\Omega
   \end{equation}
   integrated over the domain $\Omega$.
   Such a spectrum is characteristic of quasiperiodic dynamics, as recently observed for a similar flow by \cite{jfm:leclerc:2019}.
   As demonstrated below, the two main frequencies correspond either to the dynamics of the vortical structures along the shear layer ($\omega_s$) or to the low-frequency unsteadiness taking place within the cavity ($\omega_c$).
   The power spectrum consists of approximately discrete peaks, each of which can be accounted for by the sum or difference of these fundamental frequencies and their harmonics.
   The observation that this spectrum can be generated using only two main frequencies lets us hypothesize that the dynamics of the fluctuation $\bm{u}^{\prime}(\bm{x}, t)$ around the mean flow $\bar{\bm{u}}(\bm{x})$ are amenable to a low-dimensional representation.
   %As detailed in Section~\ref{sec: correlations}, it is this feature which allows us to identify a compact representation of the flow using polynomial correlations.
   %The rest of this work is thus dedicated to the identification of this low-order model by means of statistics, dimensionality reduction and machine learning.
   
%   \begin{figure}
%      \centering
%      \includegraphics[width=.66\textwidth]{instantaneous_vorticity_field}
%      \caption{Instantaneous vorticity field for the shear-driven cavity flow at $Re=7500$ highlighting the vortical structures developing along the shear layer.}
%      \label{fig: plant description -- instantaneous visualization}
%   \end{figure}

%%%%%%%%%%%%%%%%%%%%%%%%%%%%%%%%%%%%%
%%%%%     NUMERICAL METHODS     %%%%%
%%%%%%%%%%%%%%%%%%%%%%%%%%%%%%%%%%%%%
\section{Modal analysis}
\label{sec: modes}

Linear modal analysis is a powerful tool for extracting low-dimensional coherent structure in flows, even those characterized by strong nonlinearity.
Here we give only a brief description; see~\citet{aiaa:taira:2017} for a comprehensive survey.
We focus on truncated (rank $r$) affine space-time decompositions of the form
  \begin{equation}
    \bm{u}(\bm{x}, t) \simeq \bar{\bm{u}}(\bm{x}) + \sum_{k=1}^r \bm{\psi}_k(\bm{x}) a_k(t),
    \label{eq: modes -- expansion}
  \end{equation}
including for example global stability analysis~\citep{Theofilis2011arfm}, proper orthogonal decomposition~\citep{atrwp:lumley:1967, book:holmes:1996}, and dynamic mode decomposition~\citep{jfm:schmid:2010,jfm:rowley:2009}, but excluding approaches such as nonmodal stability analysis~\citep{Schmid2007arfm, McKeon2010jfm} and spectral proper orthogonal decomposition~\citep{jfm:towne:2018}.
Broadly speaking, the goal of modal analysis is to identify a suitable basis $\{ \bm \psi_k \}_{k=1}^r$ in which to represent the flow kinematics, while the reduced-order dynamical systems models discussed in Section~\ref{sec: rom} treat the time evolution of the coefficients $\bm a(t)$.
Since the state is specified by the $r$-dimensional coefficient vector, equation~\eqref{eq: modes -- expansion} is a linear dimensionality reduction.

\subsection{Proper orthogonal decomposition}
\label{sec: modes -- pod}

One of the most widely used techniques for dimensionality reduction and modal analysis is proper orthogonal decomposition (POD), which solves the optimization problem
  \begin{equation}
  \label{eq: modes -- pod-opt}
    \begin{gathered}
      \minimize_{\{ \bm \psi_k \} }  \left \langle \norm{ \bm u'  - \sum_{k=1}^r \bm \psi_k  \ip{\bm u' }{\bm \psi_k } }^2 \right \rangle  \qquad
      \subjectto  \ip{\bm \psi_j}{\bm \psi_k} = \delta_{jk}
    \end{gathered}
  \end{equation}
  in the norm induced by the energy inner product
  \begin{equation}
      \ip{\bm u}{\bm v} \equiv \int_\Omega \bm u(\bm x) \cdot \bm v^*(\bm x) ~ \dd \Omega,
  \end{equation}
  where $\langle \cdot \rangle$ is an ensemble average, approximated in practice by a time average, and $\delta_{jk}$ is the Kroenecker delta.
  For data on a nonuniform mesh, the inner product is computed with a weighted Riemann sum,  approximating $\dd \Omega$ with the mass matrix of the discretization.
  Thus, the objective is to minimize the residual energy in a linear subspace of $r$ orthonormal modes, providing an optimal low-rank representation of the flow.
  
%   \begin{figure}
%     \centering
%     \subfigure{\includegraphics[width=.3\textwidth]{dimensionality_reduction}} \hspace{1cm}
%     \subfigure{\includegraphics[width=.4\textwidth]{svd}}
%     \caption{Schematic representation of the low-rank approximation of the data matrix $\mathbfcal{X}$ by means of singular value decomposition. Each column of $\mathbfcal{X}$ contains one snapshot obtained from direct numerical simulation. The matrix $\mathbfcal{U}$ contains the space-dependent POD modes while $\mathbfcal{V}$ contains the associated temporal evolutions. Finally, the diagonal matrix $\boldsymbol{\Upsigma}$ contains the singular values whose square characterizes the amount of variance explained by the associated singular pairs.}
%     \label{fig: modes -- singular value decomposition}
%   \end{figure}

  This problem can be solved with the calculus of variations, leading to the result that the modes $\{ \bm \psi_k \}$ are eigenfunctions of the correlation tensor $\mathbfcal{C}(\bm x, \bm x')$:
  \begin{equation}
  \label{eq: modes -- pod-eigenproblem}
      \int_\Omega \mathbfcal{C}(\bm x, \bm x') \bm \psi_k(\bm x') \dd \Omega' = \sigma_k^2 \bm \psi_k(\bm x),
  \end{equation}
  where $\mathbfcal{C}(\bm x, \bm x') = \left \langle \bm u(\bm x, t) \bm u^*(\bm x', t) \right \rangle$ and $\{ \sigma_k \}$ are the POD eigenvalues, representing the average fluctuation kinetic energy captured by each mode.
  The coefficients $\bm a(t)$ can be extracted with the projection $a_k = \ip{\bm u'}{\bm \psi_k}$.
  In practice the correlation tensor is often not feasible to construct, since it scales with the square of the discretized state dimension.
  Instead it is approximated numerically with either a singular value decomposition (SVD) or the snapshot method%, as illustrated in Figure~\ref{fig: modes -- singular value decomposition}
  ~\citep{qam:sirovich:1987, book:holmes:1996, aiaa:taira:2017}.
  In this work we use the latter, since it does not require storing the entire time series of high-dimensional discretized velocity fields in memory.
  
  The method of snapshots is based on simple linear algebraic manipulations of the discretized form of the eigenvalue problem~\eqref{eq: modes -- pod-eigenproblem}.
  We omit a derivation here as it is given in standard references, e.g.~\citet{book:holmes:1996}.
  Rather than form the spatial correlation tensor $\mathbfcal{C}(\bm x, \bm x')$, we compute a temporal correlation matrix $\mathsfbi{R}$ with entries defined by
  \begin{equation}
      R_{jk} = \frac{1}{M} \ip{\bm u(t_j)}{\bm u(t_k)}, \hspace{1cm} j, k = 1, 2, \dots, M.
  \end{equation}
  The temporal correlation matrix $\mathsfbi{R}$ has dimensions $M \times M$, and is typically much smaller than the discretized spatial correlation tensor.
  The eigenvalues of $\mathsfbi{R}$ approximate those of $\mathbfcal{C}$, and the modes that solve the discretized form of~\eqref{eq: modes -- pod-eigenproblem} are also given by
  \begin{equation}
      \bm \psi_k = \frac{1}{\sigma_k \sqrt{M}} \sum_{j=1}^M \bm u(t_j) W_{jk},
  \end{equation}
  where $\mathsfbi{R}\mathsfbi{W} = \mathsfbi{W} \boldsymbol{\Sigma}^2$ is the eigendecomposition of $\mathsfbi{R}$.
  
  The proper orthogonal decomposition has the following useful properties:
  \begin{enumerate}
      \item The spatial modes form an orthonormal set: $\ip{\bm \psi_j}{\bm \psi_k} = \delta_{jk}$.
      \item The temporal coefficients are linearly uncorrelated: $\left \langle a_j a_k \right \rangle = \sigma_k^2 \delta_{jk}$.
      \item The modes can be ranked hierarchically by average energy content $\sigma_k^2$.
  \end{enumerate}
  Since POD can be viewed as a continuous form of the SVD, these properties are analogous to unitarity of the matrices of left and right singular vectors.  The singular values quantify the statistical variance captured by the low-rank SVD approximation.
  As a consequence of the hierarchical ordering, the POD can be computed without \textit{a priori} specification of the rank $r$, with the truncation determined later by a threshold based on the residual energy.
  Finally, since the modes are a linear combination of DNS snapshots, the reconstruction~\eqref{eq: modes -- expansion} automatically satisfies the incompressibility constraint and boundary conditions.
  
\begin{figure}
    \centering
    \includegraphics[width=0.7\textwidth]{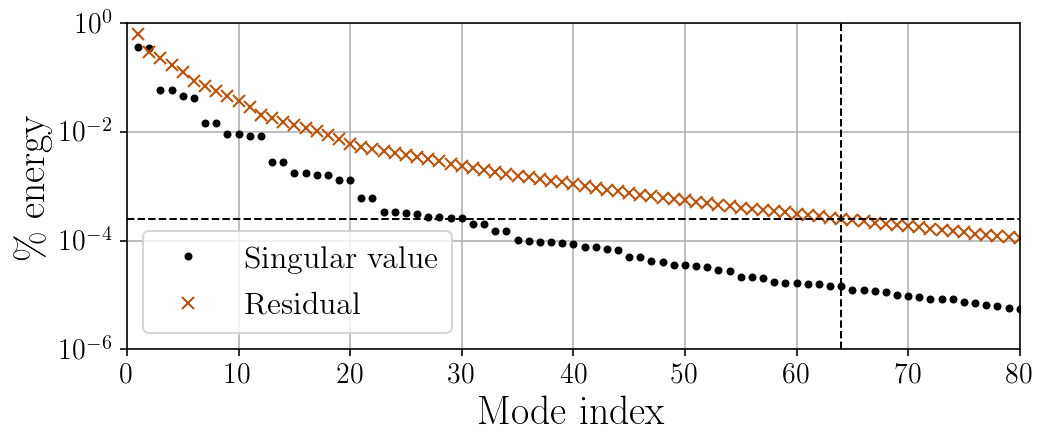}
    \vspace{-.05in}
    \caption{
    Singular value spectrum of the quasiperiodic cavity flow.
    Black dots represent the normalized squared singular values of the snapshot correlation matrix, indicating the fraction of fluctuation kinetic energy resolved by each mode.
    Red crosses indicate the fraction of residual energy, or normalized cumulative sum of squared singular values.
    Dashed lines indicate the number of modes retained ($r=64$).}
    \label{fig: modes -- singular values distribution}
        \vspace{-.05in}
  \end{figure}

  We compute the POD from 4000 fields sampled at $\Delta t = 0.05$, approximately ten times the shear layer frequency, using the method of snapshots~\citep{qam:sirovich:1987}.
  The singular value spectrum and residual energy are shown in Figure~\ref{fig: modes -- singular values distribution}.
  The singular values converge relatively quickly; the first pair of modes contain 70\% of the fluctuation kinetic energy, the first six account for $\sim 90\%$, and by $r=64$ approximately $99.97\%$ of the energy is recovered.
  We retain 64 modes for further analysis and note that our modeling results are insensitive to moderate changes in truncation.

\begin{figure}
    \centering
    \includegraphics[width=.99\textwidth]{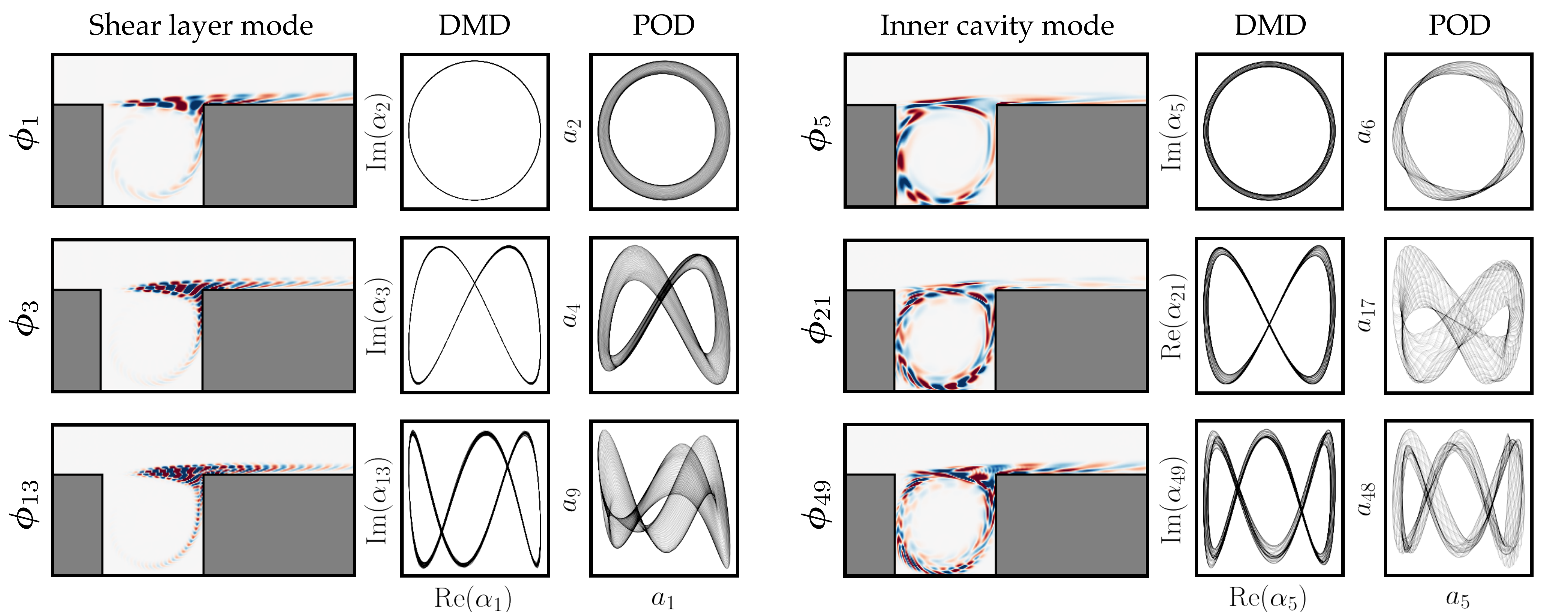}
    \vspace{-.05in}
    \caption{
      Harmonic modes identified from POD and DMD analysis.
      The spatial fields and phase portraits both indicate that certain mode pairs are harmonics arising from the description of wavelike motion in the shear layer and inner cavity.
      Because DMD is based on both spatial and temporal correlation, this structure is especially pronounced in the DMD coefficients.
      The vorticity plots are real parts of the DMD modes, but analogous modes exist in the POD basis.
    }
    \label{fig: modes -- harmonics}
    \vspace{-.04in}
    \end{figure}

  Still, as we will show in Section~\ref{sec: correlations}, the intrinsic dimensionality of the system is much smaller than that of the linear subspace required for reconstruction.
  As with the advection system in Section~\ref{sec: example}, this is partly due to the representation of traveling waves, as shown in Figure~\ref{fig: modes -- harmonics}.
  This is made more clear by a dynamic mode decomposition analysis.

\subsection{Dynamic mode decomposition}
\label{sec: modes -- dmd}

Although proper orthogonal decomposition is guaranteed to provide an energy-optimal spatial reconstruction of the flow field, it sacrifices all temporal information in the computation of the correlation tensor.
The POD basis is therefore purely kinematic and contains no dynamic information.
An alternative approach is to compute the discrete Fourier transform of the fields, which suffers from the opposite issue: frequency information is perfectly resolved, but the result is not necessarily associated with a useful reduced-order linear subspace for kinematic representation.
Dynamic mode decomposition (DMD), introduced by~\citet{jfm:schmid:2010}, is a useful compromise between these extremes.

DMD seeks to approximate a discrete-time linear evolution operator defined by
\begin{equation}
  \label{eq: modes -- dmd-opt}
    \minimize_{\mathbfcal{A}} \left \langle \norm{ \bm u(\bm x, t_{n+1}) - \mathbfcal{A} \bm u(\bm x, t_{n}) }^2 \right \rangle.
\end{equation}
The description of nonlinear dynamics in terms a linear evolution operator acting on observables has a deep connection to Koopman theory~\citep{jfm:rowley:2009, arfm:mezic:2013, Brunton2021koopman}.
We will discuss this in Section~\ref{sec: discussion}, but in terms of modal analysis it is more useful to think of DMD as solving an alternative optimization to equation~\eqref{eq: modes -- pod-opt}.

Given a series of snapshots, a least-squares solution to equation~\eqref{eq: modes -- dmd-opt} could be found in terms of the pseudoinverse of the snapshot matrix.
However, this calculation is typically computationally prohibitive, ill-conditioned, and disregards low-dimensional structure in the flow.
Instead, DMD seeks to approximate the spectral properties of the operator $\mathbfcal{A}$ without explicitly forming it.
There are a variety of algorithms to compute DMD in conditions with limited or noisy data, but beginning with high-fidelity DNS snapshots we follow the simple \textit{exact DMD} algorithm introduced by~\citet{jcd:tu:2014}.

Beginning with a truncated POD basis and associated coefficients $\{ \bm a(t_n) \}_{n=1}^M$, the coefficients are arranged into time-shifted matrices $\mathsfbi{X} = \begin{bmatrix} \bm a(t_1) & \bm a(t_2) & \cdots & \bm a(t_M-1) \end{bmatrix}$ and $\mathsfbi{X}^\prime = \begin{bmatrix} \bm a(t_2) & \bm a(t_3) & \cdots & \bm a(t_M) \end{bmatrix}$.
Here we assume the $\{t_n\}$ are evenly sampled in time, but it is possible to account for situations where this is not the case.
A least-squares solution to~\eqref{eq: modes -- dmd-opt} in the POD subspace is $ \tilde{\mathsfbi{A}} = \mathsfbi{X}^\prime \mathsfbi{X}^+$, where $\mathsfbi{X}^+$ is the pseudoinverse.
With some assumptions, the spectrum of $\mathbfcal{A}$ is approximated by the spectrum of $\tilde{\mathsfbi{A}}$, which can now be easily computed via an eigendecomposition:
\begin{equation}
    \tilde{\mathsfbi{A}} = \mathsfbi{V} \bm{\Lambda} \mathsfbi{V}^{-1}.
\end{equation}
For details on theory and algorithms of dynamic mode decomposition, see~\citet{jcd:tu:2014, book:kutz:2016}.
Based on this eigendecomposition, complex-valued DMD modes $\{ \bm{\phi}_k(\bm x) \}$ and associated projection coefficients $\bm \alpha(t)$ are linear combinations of the POD modes and coefficients, given by
\begin{equation}
    \bm \phi_k(\bm x) = \sum_{j=1}^r \bm \psi_j(\bm x) V_{jk} \hspace{1.5cm} \bm \alpha(t) = \mathsfbi{V}^{-1} \bm a(t).
\end{equation}
In principle the approximate time evolution is specified by the DMD eigenvalues $\{\lambda_k\} = \mathrm{diag}(\bm \Lambda)$, but in terms of reduced-order modeling the decomposition can also be viewed as an alternative expansion to~\eqref{eq: modes -- expansion}:
\begin{equation}
\label{eq: modes -- dmd-expansion}
    \bm u(\bm x, t) \simeq \bar{\bm u}(\bm x) + \sum_{k=1}^r \bm \phi_k (\bm x) \alpha_k(t).
\end{equation}
The DMD coefficient vector $\bm \alpha(t)$ may then be modeled as a time series, as with the POD coefficients $\bm a(t)$.
  
  \begin{figure}
    \centering
    \includegraphics[width=0.95\textwidth]{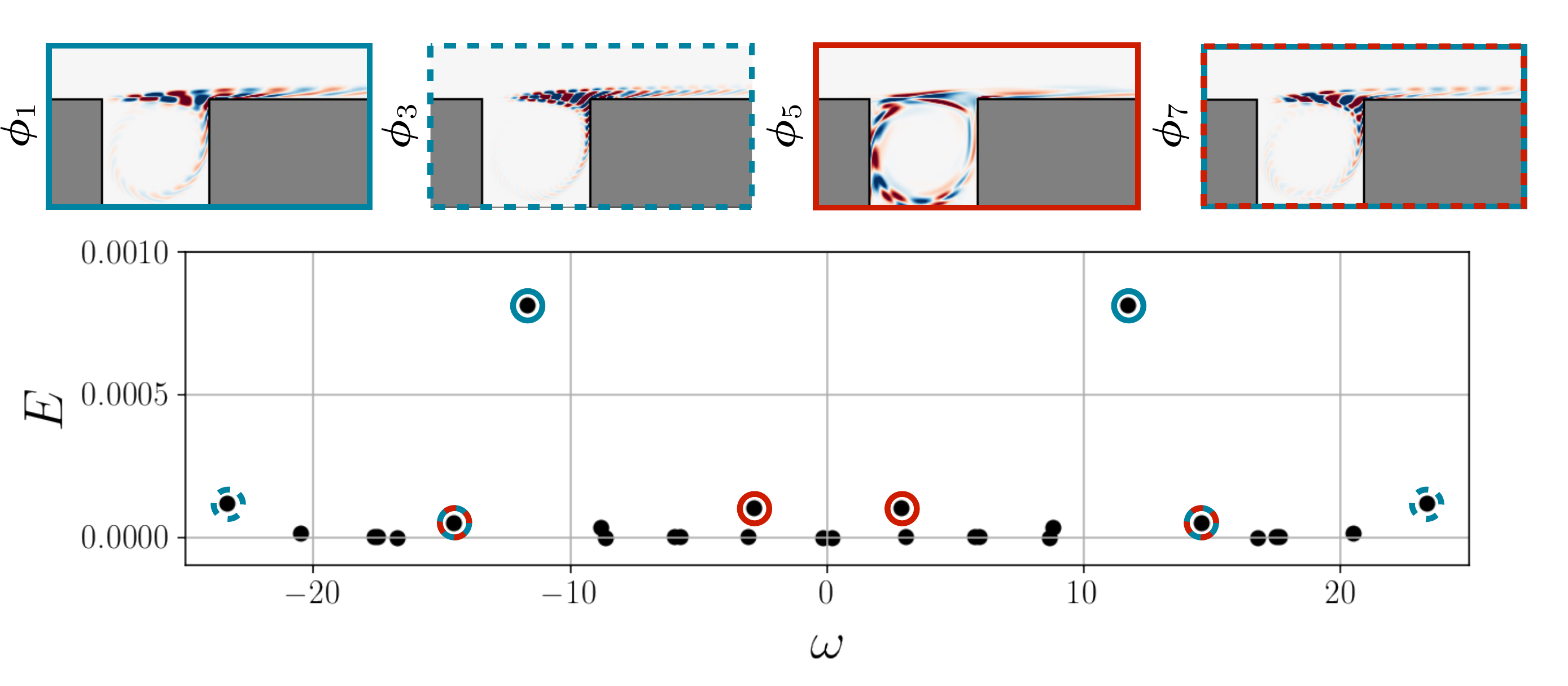}
    \caption{
    DMD frequencies $\omega_k$ and average energy $E_k = \langle |\alpha_k|^2 \rangle$ along with  vorticity plots for the real part of the most energetic modes.
    The second mode pair ($k=3, 4$) is a harmonic of the leading pair ($k=1,2$), while the third pair ($k=5, 6$) represents the low-frequency inner cavity motion.
    Other modes (e.g. $k=7, 8$) are either harmonics or indicate nonlinear frequency crosstalk between these leading modes, as in Figure~\ref{fig: plant description -- fourier spectrum}.}
    \label{fig: modes -- dmd-spectrum}
  \end{figure}

% One drawback to the DMD representation is that there is no natural ordering of the modes.
% Some possible intuitive criteria are to sort modes by either eigenvalues (descending $\Re(\lambda)$), mean energy content (descending $\langle \alpha^2(t) \rangle$), or frequency (increasing $\Im(\lambda)$).
% In this case we do not further truncate the basis beyond the original POD dimensionality of $r=64$, so this is only a matter of indexing and we choose to sort by average energy content.
% However, in general the correct choice is not necessarily clear and to our knowledge this remains an open problem.

This representation is essentially a similarity transformation of the POD basis; the two encode the same information and span the same subspace.
However, the time-dependence in the optimization problem leads DMD to transform the POD basis to modes that tend to have similar frequency content.
In terms of the present analysis, Figure~\ref{fig: modes -- harmonics} illustrates the practical relevance of this.
Whereas POD happens to identify modes that are roughly coherent in time by coincidence only, the DMD modes are closer to pure harmonics.
This perspective on DMD also explains the approximately discrete peaks in Figure~\ref{fig: plant description -- fourier spectrum}; each DMD eigenvalue (shown in Figure~\ref{fig: modes -- dmd-spectrum}) can be identified with some combination of the fundamental frequencies of modes $k=1$ and $k=5$.

Based on the results in Sections~\ref{sec: correlations} and~\ref{sec: results}, we hypothesize that DMD also filters frequency content and accentuates nonlinear correlations for modes that are not pure harmonics.
This approximate nonlinear algebraic dependence clearly indicates a manifold structure of much lower dimensionality than the linear subspace.
Given these implications, the results presented below are based on the DMD expansion~\eqref{eq: modes -- dmd-expansion}.

%%%%%%%%%%%%%%%%%%%%%%%%%%%
%%%%%     RESULTS     %%%%%
%%%%%%%%%%%%%%%%%%%%%%%%%%%
\section{Reduced-order models}
\label{sec: rom}

The modal analyses discussed in Section~\ref{sec: modes} may be viewed as linear dimensionality reduction methods that transform the system to a compact coordinate system in which low-dimensional dynamical systems models can be developed.
In addition to an inexpensive surrogate for the flow, such models can provide valuable insight into latent structure of the physical solutions.
Broadly speaking, two of the most common approaches to nonlinear reduced-order modeling are projection-based models and data-driven system identification, though many more tools are available for linear model reduction; see for instance~\citet{Antoulas2005, Benner2015siam}.
In this section we give a brief overview of relevant material on projection-based modeling (Section~\ref{sec: rom -- galerkin}) and the SINDy framework for system identification (Section~\ref{sec: rom -- sindy}).

%   This section briefly introduces the reader to the basic mathematical concepts used in the present work for low-dimensional system identification.
%   First, attention is focused on dimensionality reduction using \emph{proper orthogonal decomposition}, a necessary pre-processing step of the high-dimensional dataset obtained from direct numerical simulation.
%   Then, the system identification procedure is described.
%   It falls within the framework of \emph{sparse identification of nonlinear dynamics} (SINDy) recently proposed in \cite{pnas:brunton:2016}.
%   Additionally, a brief overview of POD-Galerkin projection is also given as it is currently state-of-the-art approach for reduced-order modeling.

%%%%%%%%%%%%%%%%%%%%%%%%%%%%%%%%%%%%%%%%%%%
%%%%%     POD-GALERKIN PROJECTION     %%%%%
%%%%%%%%%%%%%%%%%%%%%%%%%%%%%%%%%%%%%%%%%%%
\subsection{POD-Galerkin modeling}
\label{sec: rom -- galerkin}
  
In projection-based modeling, the discretized governing equations are projected onto an appropriate modal basis.
For simple geometries, this might be done analytically, as for the periodic problems in Section~\ref{sec: example} and in~\citet{jfm:noack:1994}, for instance.
Although general and expressive, this approach becomes challenging on complex domains and does not take advantage of structure in the solutions to the particular PDE.
As a result, it is increasingly common to project onto an empirical basis, such as POD modes.
Assuming that the flow is statistically stationary and the ensemble is sufficiently resolved, this provides optimal kinematic resolution in an orthonormal basis.
The following Galerkin projection procedure then leads to a minimum-residual system of ODEs in this basis.

Let $\mathbfcal{N}[\bm u] = 0$ be the Navier-Stokes equations in implicit form.
By approximating the flow field with a truncated linear combination of basis functions as in~\eqref{eq: modes -- expansion}, we expect some residual error $\bm r(x, t)$ in the approximated dynamics defined by
\begin{equation}
    \label{eq: rom -- NS-residual}
    \bm r(\bm x, t) = \mathbfcal{N}\left[ \bar{\bm u}(\bm x) + \sum_{\ell = 1}^r \bm \psi_\ell(\bm x) a_\ell(t) \right].
\end{equation}

In order to minimize the residual in this basis, the Galerkin projection condition is that that the residual be orthogonal to each mode:
\begin{equation}
    0 = \ip{\bm r}{\bm \psi_k} = \ip{\mathbfcal{N}\left[ \bar{\bm u}(\bm x) + \sum_{\ell = 1}^r \bm \psi_\ell(\bm x) a_\ell(t) \right]}{\bm \psi_k}.
\end{equation}
This leads to the linear-quadratic system of ODEs~\citep{book:holmes:1996, book:noack:2011}
\begin{equation}
    \label{eq: rom -- galerkin}
    \dot{a}_k = C_k + \sum_{\ell = 1}^r L_{k \ell} a_\ell + \sum_{\ell = 1}^r \sum_{m = 1}^r Q_{k \ell m} a_\ell a_m,
\end{equation}
with constant, linear, and quadratic terms given by
\begin{subequations}
\begin{align}
    C_k &= \ip{-\nabla \cdot (\bar{\bm u} \otimes \bar{\bm u} ) + \frac{1}{\Rey} \nabla^2 \bar{\bm u} }{\bm \psi_k} \\
    L_{k \ell} &= \ip{-\nabla \cdot (\bar{\bm u} \otimes \bm \psi_\ell +  \bm \psi_\ell \otimes \bar{\bm u } ) + \frac{1}{\Rey} \nabla^2 \bm \psi_\ell }{\bm \psi_k} \\
    Q_{k \ell m} &= \ip{ -\nabla \cdot (\bm \psi_m \otimes \bm \psi_\ell +  \bm \psi_\ell \otimes \bm \psi_m ) } {\bm \psi_k}.
\end{align}
\end{subequations}
Note that the constant term vanishes if the flow is expanded about a steady-state solution of the governing equations.
Since the mean flow in this case is not a solution, this term represents important mean-flow forcing and is not negligible.
Here we have also neglected the pressure term, though including it does not significantly change any of the results; for detailed discussion of this point see~\citet{jfm:noack:2005}.

In principle, we might expect that the POD-Galerkin system~\eqref{eq: rom -- galerkin} leads to approximate solutions with comparable accuracy to the resolution of the expansion basis.
However, for reasons introduced in Section~\ref{sec: example}, the long-time behavior of the reduced-order model may deviate significantly from that of the underlying physical system.
In particular, solutions of the model are not constrained to lie on an invariant manifold of the flow.
For instance, coefficients associated with shear layer or inner-cavity harmonics evolve independently from the fundamental modes, eventually leading to an unphysical loss of coherence.

\begin{figure}
\centering
\includegraphics[width=\textwidth]{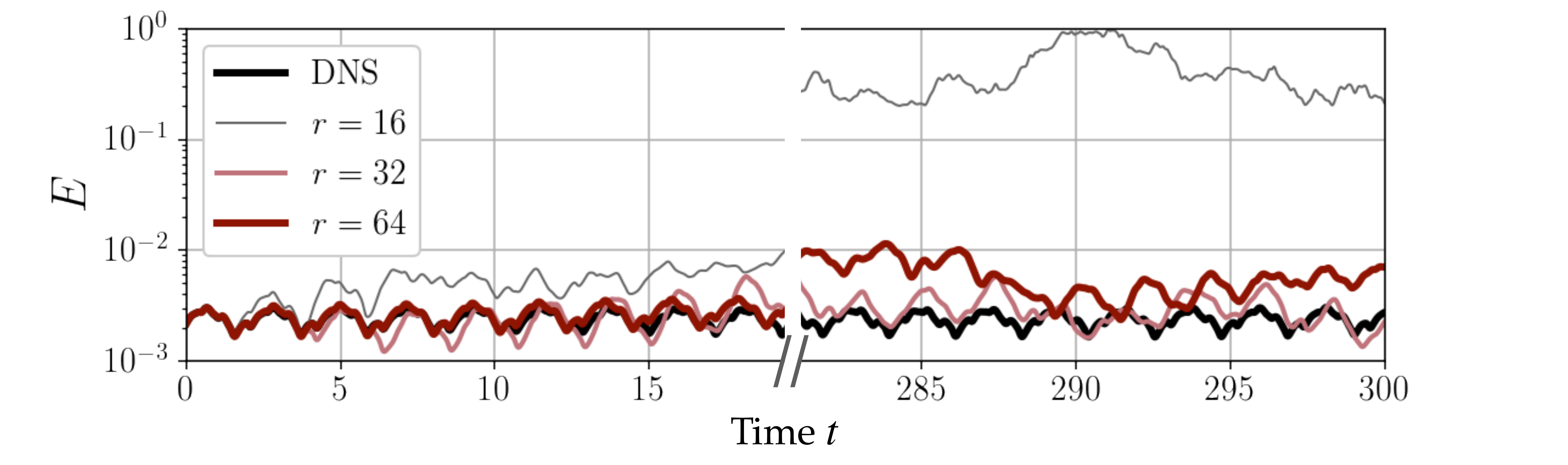}
\vspace{-.2in}
\caption{Evolution of the fluctuation kinetic energy predicted by POD-Galerkin reduced-order models of various dimensions along with DNS values.
Though all values of $r$ shown here capture sufficient dissipation to remain at finite energy, none resolves the true quasiperiodic dynamics.}
\label{fig: rom -- pod-galerkin energy evolution}
\end{figure}

Figure~\ref{fig: rom -- pod-galerkin energy evolution} shows the evolution of the fluctuation kinetic energy as predicted by the POD-Galerkin system for various levels of truncation $r$.
Although the estimate does tend to improve with increasing $r$, none of these models capture the quasiperiodic dynamics of the flow, and most exhibit significant instability.
This is true despite (and, we will argue, because of) the fact that these models have many more kinematic degrees of freedom than the true dynamics underlying the post-transient cavity flow.

Finally, a model similar to~\eqref{eq: rom -- galerkin} may also be derived beginning with the DMD expansion~\eqref{eq: modes -- dmd-expansion}.
In this case, since the DMD basis is not orthonormal, the Galerkin projection must be replaced with the more general oblique projection of Petrov-Galerkin methods~\citep{Benner2015siam}.
We implement this with a coordinate transform of the standard POD-Galerkin model based on the DMD eigenvector matrix $\mathsfbi{V}$, i.e. $\bm a = \mathsfbi{V} \bm \alpha $.

Since this transformation is a rotation in the space of modal coefficients, the dynamics and qualitative behavior of the DMD-Galerkin models do not change compared to Figure~\ref{fig: rom -- pod-galerkin energy evolution}.
However, in the following we exploit nonlinear correlations in the coefficients to restrict the dynamics to the manifold of the flow; this is more convenient in the near-harmonic DMD basis, shown in Figure~\ref{fig: modes -- harmonics}.
Since DMD analysis seeks to approximate the spectrum associated with a linear evolution operator of the flow, this might be considered analogous to the diagonalization step of a center manifold or normal form analysis.

%%%%%%%%%%%%%%%%%%%%%%%%%%%%%%%%%%%%%%%%%
%%%%%     SYSTEM IDENTIFICATION     %%%%%
%%%%%%%%%%%%%%%%%%%%%%%%%%%%%%%%%%%%%%%%%
\subsection{Sparse identification of nonlinear dynamics}
\label{sec: rom -- sindy}

As an alternative to projection-based reduced-order modeling, a low-dimensional system can be approximated directly from the data in a procedure typically called system identification.
In a continuous-time setting, this is done by estimating parameters $\bm \Xi$ for a function $\bm f(\bm a; \bm \Xi) \approx \dot{\bm a} $ that solve the approximation problem
\begin{equation}
\label{eq: rom -- objective}
    \minimize_{\bm \Xi} \left \langle \norm{ \dot{\bm a} - \bm f(\bm a; \bm \Xi) }^2 \right \rangle.
\end{equation}
This is similar to the residual minimization in Galerkin projection, except that knowledge of the governing equations is neither required nor assumed.
Instead, some intuition about the structure of the dynamics is typically encoded in the parameterization of  $\bm f$.
Different parameterizations lead to symbolic regression~\citep{science:schmidt:2009}, operator inference~\citep{Peherstorfer2016cmame}, or deep learning~\citep{prsa:vlachas:2018}.
In the discrete-time counterpart to~\eqref{eq: rom -- objective}, the NARMAX framework provides a powerful approach that can incorporate time delays, stochastic forcing, and exogenous inputs~\citep{book:billings:2013}.

In this work we apply the \emph{sparse identification of nonlinear dynamics} (SINDy) approach to system identification~\citep{pnas:brunton:2016}.
Let $\boldsymbol{\Uptheta}(\bm a)$ denote a library of candidate functions of the time series $\bm a(t)$, e.g.
\begin{equation}
\label{eq: rom -- sindy-lib}
    \boldsymbol{\Uptheta}\left(\begin{bmatrix}
    a_1 & a_2 
    \end{bmatrix}^T \right) = \begin{bmatrix}
    a_1 & a_2 & a_1^2 & a_1 a_2 & a_2^2 & \cdots
    \end{bmatrix}^T.
\end{equation}
% \begin{equation}
% \label{eq: rom -- sindy-lib}
%     \boldsymbol{\Uptheta}\left(\begin{bmatrix}
%     | & | \\
%     a_1 & a_2 \\
%     | & |
%     \end{bmatrix}^T \right) = \begin{bmatrix}
%     | & | & | & | & | & & \\
%     a_1 & a_2 & a_1^2 & a_1 a_2 & a_2^2 & \cdots \\
%     | & | & | & | & | & &
%     \end{bmatrix}^T.
% \end{equation}
We seek a sparse approximation $\dot{\bm a}\approx \bm f(\bm a; \bm \Xi) = \bm \Xi \boldsymbol{\Uptheta}(\bm a) $ in the range of these candidate functions.
We can frame this as a linear algebra problem by forming the $r \times M$ data matrix $\mathsfbi{X}$ as in Section~\ref{sec: modes -- dmd}, where each column is a snapshot of modal coefficients in time.
Similarly, we estimate the time derivative $\dot{\mathsfbi{X}}$, in this case with second-order central differences.
The SINDy formulation of the optimization problem~\eqref{eq: rom -- objective} is then
\begin{equation}
\label{eq: rom -- sindy-objective}
    \minimize_{\bm \Xi} \norm{ \dot{\mathsfbi{X}} - \bm \Xi \boldsymbol{\Uptheta}(\mathsfbi{X})  }_2^2 + \gamma \norm{\bm \Xi}_0,
\end{equation}
where $\norm{\cdot}_p$ indicates the $p-$norm and $\gamma$ is some regularization weight.
This optimization problem is non-convex and requires a combinatorial search over function combinations.
To avoid this, we follow~\citet{Loiseau2019tcfd} and approximate the solution to~\eqref{eq: rom -- sindy-objective} with the greedy forward regression orthogonal least squares (FROLS) algorithm used in NARMAX analysis~\citep{book:billings:2013}.
See~\citet{pnas:brunton:2016, jfm:loiseau:2018a, Loiseau2019tcfd} for details on SINDy reduced-order modeling of fluid flows.

Although the systems obtained from POD-Galerkin projection will be dense in general, we expect that the dynamics can be closely approximated by a sparse combination of candidate functions.
This may be justified intuitively by the sparse structure of the triadic interactions in isotropic flow (see Section~\ref{sec: example -- burgers}), where only $r^2$ out of $r^3$ possible interactions are admissible.
Moreover, as observed in Section~\ref{sec: modes -- dmd}, DMD approximates a diagonalization of the evolution operator, so that by analogy with normal form theory we may reasonably hope for a minimal representation of the dynamics if we work with DMD coefficients $\bm \alpha$ rather than POD coefficients $\bm a$.
More generally, sparsity promotion reflects the inductive bias of Occam's razor or Pareto analysis, where we expect that the most important features of the dynamics will be due to a small subset of terms.

For incompressible flows, it has been repeatedly demonstrated that a library of low-order polynomials provides a good basis of functions.
Quadratic terms are clearly necessary to capture the advective nonlinearity of the Navier-Stokes equations, but cubic terms allow the model to resolve Stuart-Landau-type nonlinear stability mechanisms~\citep{jfm:loiseau:2018a}.
From a dynamical systems perspective, higher-order terms may be necessary to describe phenomena such as subcritical bifurcations, but are not necessary to resolve the nonlinear oscillator behavior in the present case.

For PDEs with more general nonlinearity, low-order polynomials still present an attractive basis for SINDy.
In many interesting regimes the effect of the nonlinearity may be relatively weak, so that quadratic and cubic polynomials can be seen as second- or third-order Taylor series approximations to the underyling nonlinearity.
Moreover, even strongly nonlinear systems can be \emph{lifted} with a change of variables to a coordinate system wherein the dynamics are linear-quadratic~\citep{Rowley2004pd, Qian2020}.
% The SINDy approach is therefore a simple and powerful tool for data-driven reduced-order modeling.

%%%%%%%%%%%%%%%%%%%%%%%%%%%%%%
%%%%%     DISCUSSION     %%%%%
%%%%%%%%%%%%%%%%%%%%%%%%%%%%%%
\section{Nonlinear correlations}
\label{sec: correlations}

Sections~\ref{sec: modes} and~\ref{sec: rom} recapitulate well-known methodology for analyzing and modeling unsteady fluid flows.
With a modal expansion and model reduction, the formally infinite-dimensional PDE can be reduced to a set of coupled, nonlinear ODEs mimicking the structure of the physical system.
However, the POD analysis indicates that dozens of modes are necessary for an approximately complete kinematic representation in a linear basis, while the DMD analysis and power spectrum suggest that the dynamics of the flow are quasiperiodic.
A minimal description of the post-transient flow should therefore require only a pair of oscillators evolving on a 2-torus, comprising four degrees of freedom.

This discrepancy can be understood qualitatively in light of the discussion in Section~\ref{sec: example}.
Advection of nearly periodic fluctuations leads to the appearance of harmonic modes, as shown in Figures~\ref{fig: modes -- harmonics} and~\ref{fig: modes -- dmd-spectrum}.
Similarly, crosstalk between the incommensurate dominant frequencies gives rise to modes that are not pure harmonics of either frequency.
In the low-dimensional subspace spanned by the leading POD/DMD modes, these can be viewed as triadic interactions in the frequency domain.
Again, since seek structure that is coherent at distinct frequencies, we will focus on the DMD coefficients in the following.

\subsection{A model quasiperiodic cascade}
\label{sec: correlations -- model}

Consider a model system of ODEs including cascading triadic-type interactions of the form of~\eqref{eq: example -- burgers-galerkin} where self-sustaining oscillations drive higher-order degrees of freedom:
\begin{subequations}
\label{eq: correlations -- toy}
\begin{align}
    a_1 & = A e^{i \omega_a t}, &
    b_1 &= B e^{i \omega_b t} &&\\
    \dot{a}_k &= \sum_{|\ell| < k} a_{\ell}a_{k - \ell}, &
    \dot{b}_k &= \sum_{|\ell| < k} b_{\ell}b_{k - \ell}, &
    \dot{c}_k &= \sum_{|\ell| < k} a_{\ell}b_{k - \ell}, \qquad k > 1
\end{align}
\end{subequations}
%with symmetry $a_{-k} = a_k^*$, etc.
For $k=2$ the only interactions are at $|\ell| = 1$, generating second harmonics for $a_2$ and $b_2$, and oscillations at $\omega_a + \omega_b$ for $c_2$.
At $k=3$, the forcing includes the $k=2$ terms, generating third harmonics for $a_3$ and $b_3$ and crosstalk for $c_3$.
Up to some complex scaling the solutions take the form
\begin{subequations}
\begin{align}
    a_2 &= A^2 e^{2i\omega_a t}, &
    b_2 &= B^2 e^{2i\omega_b t}, &&
     c_2 = A B e^{i(\omega_a + \omega_b) t},\\
    a_3 &= A^3 e^{3i\omega_a t}, &
    b_3 &= B^3 e^{3i\omega_b t}, &&
    c_3 = AB^2 e^{i(\omega_a + 2 \omega_b) t} + A^2B e^{i(2\omega_a + \omega_b) t}.
\end{align}
\end{subequations}
By $k=3$ the response of the $c_k$ crosstalk variable does not represent pure frequency content.  If these represented modal coefficients we would expect the dynamics at (for instance) $\omega_a + 2\omega_b$ and $2\omega_a + \omega_b$ correspond to different spatial structures so $c_3$ and higher-order terms would be separated into distinct coefficients.
This serves to emphasize that the temporal coefficients and low-dimensional ODEs are only convenient representations of the full spatiotemporal dynamics and are not fundamental physical quantities.

A global observable such as the energy analogue $\sum_k (|a_k|^2 + |b_k|^2 + |c_k|^2)$ will have frequency content at all integer combinations of $\omega_a$ and $\omega_b$.
As with the linear advection equation in Section~\ref{sec: example -- advection}, these higher-order coefficients can be expressed as algebraic functions of the fundamental oscillators.
For instance, the crosstalk variable $c_3$ can be replaced by $a_1 b_1^2 + a_1^2 b_1$, up to scaling.
This example shows that periodic oscillation with cascading triadic interactions can generate quasiperiodic time series, point spectra similar to Figure~\ref{fig: plant description -- fourier spectrum}, and nonlinear algebraic dependence without direct interactions between the oscillators, provided the dominant oscillation frequencies are incommensurate.

\subsection{The randomized dependence coefficient}

\begin{figure}
\centering
\includegraphics[width=0.65\textwidth]{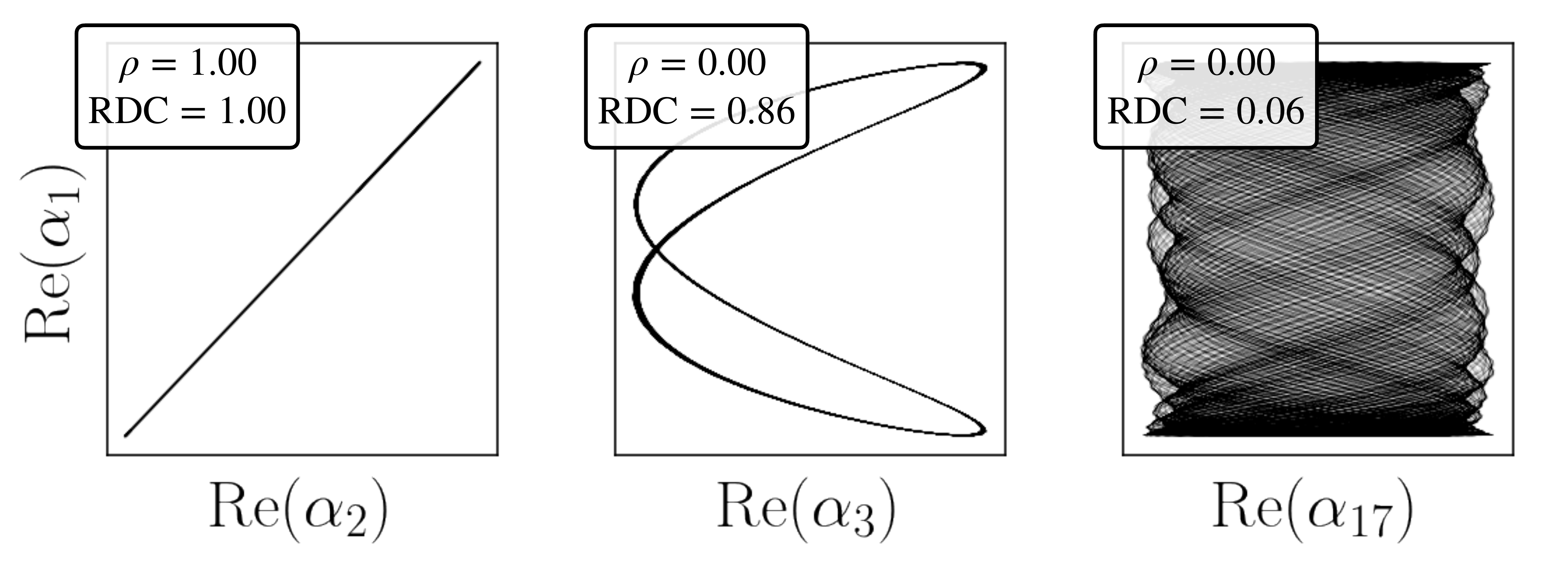}
\vspace{-.125in}
\caption{
Selected phase portraits of DMD coefficients along with measures of linear and nonlinear correlation (Pearson's $\rho$ and the randomized dependence coefficient, respectively).
While $\alpha_1$ and $\alpha_3$ are linearly uncorrelated ($\rho$ = 0), the clear functional relationship between the two is reflected in the RDC value; physically, $\alpha_3$ is a pure harmonic of $\alpha_1$.
On the other hand, $\alpha_{17}$ corresponds to a nonlinear crosstalk mode that has no clear correlation with $\alpha_1$, either linear or nonlinear.
Nevertheless, it can be accurately approximated by a simple polynomial function of the active degrees of freedom, as shown by Table~\ref{tab: results -- correlations} and Figure~\ref{fig: results -- manifold-coeff-reconstruction}.
}
\label{fig: correlations -- demo}
\vspace{-.075in}
\end{figure}

As discussed in Section~\ref{sec: modes}, the POD coefficients are guaranteed to be linearly uncorrelated.
The same is not necessarily true of DMD coefficients, but in practice they tend to be minimally correlated.
However, as in the previous example, a network of triadic interactions forced by a limited number of driving oscillators can exhibit pure algebraic dependence on the active degrees of freedom.
In other words, the higher-order variables can have perfect nonlinear correlation, even when uncorrelated in a linear sense.

This is an intuitive result, but challenging to evaluate in a principled way.
In a probabilistic setting, mutual information is the most natural metric for generalized correlation, but it requires estimating integrals over conditional probability distributions.
This is expensive and difficult for multidimensional signals, and the concept of mutual information itself is not necessarily well-suited for purely deterministic systems.
To address this issue, various nonlinear generalizations of the standard (Pearson's) linear correlation coefficient have been proposed in the statistics community.
Recently,~\citet{rdc:lopez-paz:2013} proposed the \emph{randomized dependence coefficient} (RDC) as an efficient and convenient metric for nonlinear correlation that has the properties defined by~\citet{Renyi1959} for generalized measures of dependence between variables.

\begin{figure}
\centering
\includegraphics[width=1.0\textwidth]{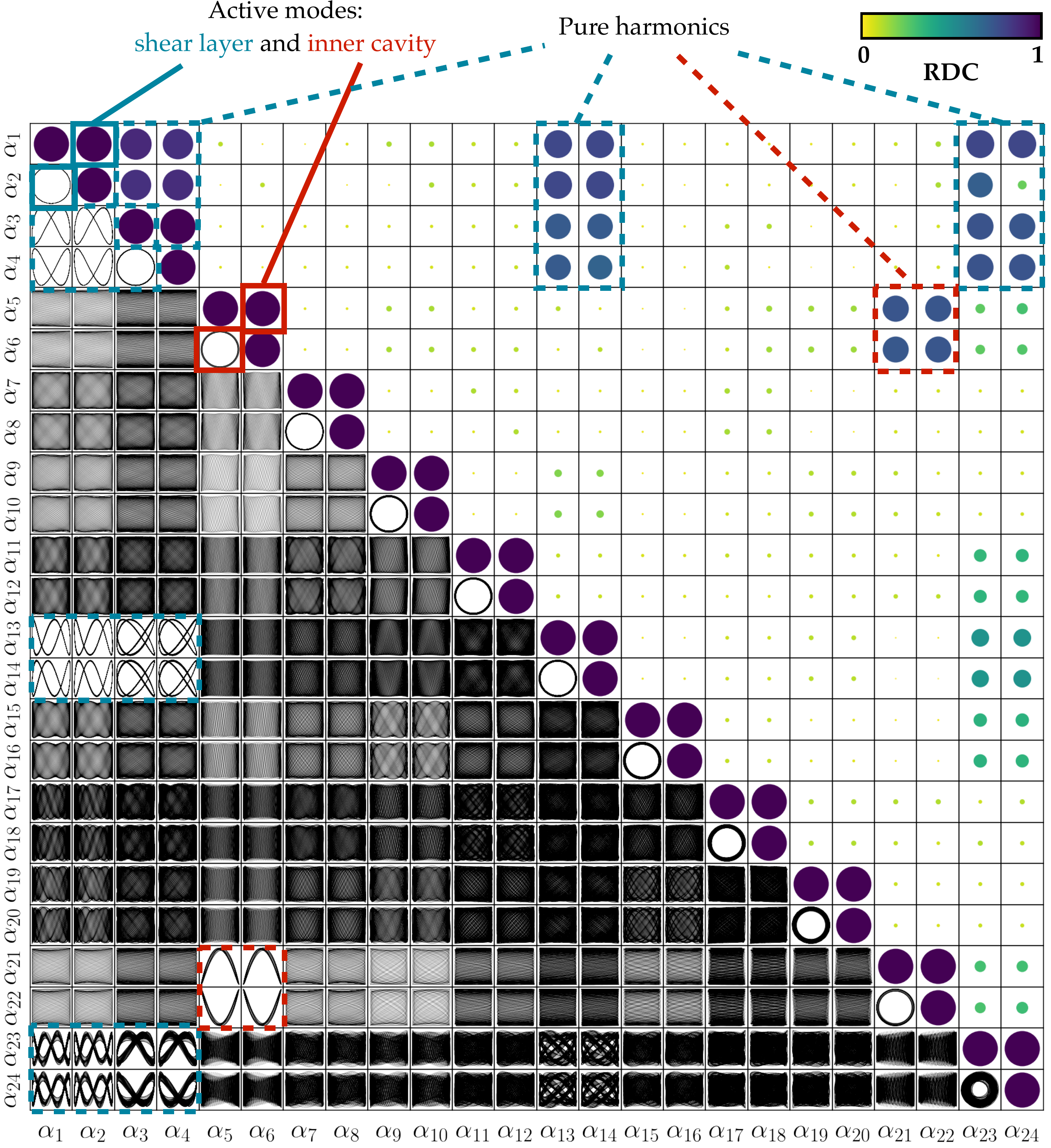}
\caption{
Identification of the active degrees of freedom with the randomized dependence coefficient (RDC).
The lower triangular portion of the figure shows phase portraits of the real (horizontal axes) and imaginary (vertical axes) DMD coefficients, while the upper triangular portion depicts the RDC values scaled linearly in color and radius.
Two approximately independent clusters can be identified: the shear layer dynamics (blue) and inner cavity oscillations (red).
Each of these is associated with a dominant mode pair (solid borders) and pure harmonics (dashed borders) that are strongly nonlinearly correlated with the dominant modes.
The other modes also have simple polynomial relationships with the active degrees of freedom but include cross terms that break the one-to-one nonlinear correlation (see Table~\ref{tab: results -- correlations} and Figures~\ref{fig: correlations -- demo} and~\ref{fig: results -- manifold-coeff-reconstruction}).
}
\label{fig: correlations -- lissajous}
\end{figure}

The RDC combines linear canonical correlations analysis with randomized nonlinear projections to estimate nonlinear dependence; details are presented in~\citet{rdc:lopez-paz:2013}.
Figure~\ref{fig: correlations -- lissajous} shows both the phase portraits of leading DMD coefficients (lower triangular portion) and the RDC values (upper triangular).
In some cases (colored outlines), the modes are clearly pure harmonics of one of the two driving mode pairs.
This is reflected in the large values of the RDC for the harmonics, indicating that these coefficients can be directly expressed as algebraic functions of one or the other driving modes.
However, based on the previous discussion we expect that coefficients representing frequency crosstalk might be multivariate functions of both driving mode pairs.
%Indeed, one of the limitations of dependence metrics such as the RDC is that it cannot predict multivariate dependence.
%It is also a metric of correlation, but contains no causal information.
In this case it is clear based on the energy content and harmonic structure of Figure~\ref{fig: correlations -- lissajous} that the pairs $(\alpha_1, \alpha_2)$ and $(\alpha_5, \alpha_6)$ are the driving degrees of freedom, but chaotic or turbulent flows might have more opaque causal structure.

\subsection{Manifold reduction via sparse regression}
\label{sec: correlations -- sparse}
Based on the RDC analysis of the previous section, it is clear that certain modes are pure algebraic functions of one or the other driving mode pairs.
In particular, harmonic modes such as those illustrated in Figure~\ref{fig: modes -- harmonics} are polynomial functions of the fundamental mode.
However, as demonstrated by the model in Section~\ref{sec: correlations -- model}, triadic interactions can also generate multivariate nonlinear correlations for modes representing frequency crosstalk.
In addition, from a dynamical systems perspective we expect that a quasiperiodic system with two dominant frequencies should have a post-transient attractor described by four degrees of freedom: two generalized amplitude and phase pairs.

If this is the case, the modes that are not pure harmonics may still be approximately polynomial functions of the shear layer and inner cavity modes.
We explore this hypothesis with the same approach as outlined in Section~\ref{sec: rom -- sindy} for the SINDy algorithm.
Denoting these ``active'' degrees of freedom by $\hat{ \bm \alpha}(t) = \begin{bmatrix}
\alpha_1 & \alpha_2 & \alpha_5 & \alpha_6
\end{bmatrix}^T$ , the library $\boldsymbol{\Uptheta}(\hat{ \bm \alpha})$ is defined as in~\eqref{eq: rom -- sindy-lib}.
We assume the full coefficient vector can be approximated as
\begin{equation}
\label{eq: correlations -- sindy}
    \bm \alpha
    \approx \mathsfbi{H} \boldsymbol{\Uptheta}
    (\hat{ \bm  \alpha}),
\end{equation}
where the coefficient matrix $\mathsfbi{H}$ is relatively sparse, as for ${\bm \Xi}$ in the SINDy optimization problem.
For rows corresponding to active degrees of freedom, $\mathsfbi{H}$ are unit vectors that produce an identity map (e.g. $\alpha_1 = \hat{\alpha}_1$)

In this case one motivation for sparse regression is that the library $\boldsymbol{\Uptheta}$ tends to be fairly ill-conditioned, so that approximation with a sparse combination of polynomials may help avoid overfitting.
In addition, based on the preceding discussions about DMD and triadic interactions in frequency space, it is reasonable to expect that relatively few combinations of the driving frequencies will correspond to each DMD coefficient.
Once the coefficient matrix $\mathsfbi{ H }$ is identified via a sparse regression algorithm (we use FROLS, as for the SINDy optimization), the functional relationships give a simple nonlinear dimensionality reduction; in this case the ``latent variables'' are the active degrees of freedom $\hat{\bm \alpha }$ and we can approximate the full coefficient vector with the function $\bm \alpha \approx \bm h (\hat{ \bm \alpha}) = \mathsfbi{H} \boldsymbol{\Uptheta}
    (\hat{\bm \alpha}) $.

The key advantage to this representation of the modal coefficients is that it dramatically restricts the dimensionality of the state space.
In the case of the quasiperiodic shear-driven cavity, we reduce from the $r=64$-dimensional subspace spanned by the DMD modes to a 4-dimensional space of $\hat{\bm \alpha}$.
Moreover, as $\alpha_2 = \alpha_1^*$ and $\alpha_6 = \alpha_5^*$, these mode pairs represent two generalized amplitude-phase pairs, as expected for dynamics with a toroidal attractor.
This eliminates the redundant variables introduced by the linear space-time decomposition via a nonlinear manifold reduction.

The benefit of this reduced state space is immediately clear for system identification; we can apply SINDy to model the evolution of $\hat{\bm \alpha}$ and reconstruct the full state from this minimal representation.
However, the manifold representation can also be used to improve the stability and accuracy of the projection-based Galerkin models.
For a general nonlinear embedding of the form $\bm a = \bm h (\hat{\bm a} )$ and dynamical system $\dot{\bm a} = \bm f (\bm a)$, consistency requires $\dot{\bm a} = \mathsfbi{J}(\hat{\bm a})\dot{\hat{\bm a}}$,
where $\mathsfbi{J}(\hat{\bm a})$ is the Jacobian of $\bm h$ evaluated at $\hat{\bm a}$.
Equivalently,
\begin{equation}
    \dv{\hat{\bm a}}{t} = \mathsfbi{J}^+(\hat{\bm a}) \dot{\bm a} = \mathsfbi{J}^+(\hat{\bm a}) \bm f( \bm h (\hat{\bm a} ) ),
\end{equation}
where $\mathsfbi{J}^+$ is the pseudoinverse of $\mathsfbi{J}$. 
This condition defines the reduced dynamics by constraining the velocity of $\bm a$ to the tangent space of the manifold defined by $\bm h$~\citep{GuckenheimerHolmes, Lee2020jcp}.

In the case of the linear-quadratic Galerkin dynamics~\eqref{eq: rom -- galerkin} with the sparse polynomial manifold equation~\eqref{eq: correlations -- sindy}, the Jacobian consists of rows of the identity matrix by virtue of the fact that the reduced states $\hat{\bm \alpha}$ are also contained in the full coefficient vector $\bm \alpha$.
Then the \emph{manifold Galerkin} dynamics are
\begin{equation}
    \label{eq: correlations -- manifold-galerkin}
    \dot{\alpha}_k = \tilde{C}_k + \sum_{\ell = 1}^r \tilde{L}_{k \ell} h_\ell( \hat{ \bm\alpha}) + \sum_{\ell = 1}^r \sum_{m = 1}^r \tilde{Q}_{k \ell m} h_\ell( \hat{ \bm\alpha}) h_m( \hat{ \bm\alpha}), \hspace{1cm} k = 1, 2, 5, 6,
\end{equation}
where the tilde denotes the original POD-Galerkin operators rotated to the DMD coordinates (see Sections~\ref{sec: modes} and~\ref{sec: rom}).
Note that if $\bm h$ contains polynomials up to order $d$, the quadratic interactions in~\eqref{eq: correlations -- manifold-galerkin} lead to an effective nonlinearity of order $2d$.
This can be viewed as a generalization of Stuart-Landau-type mean field models~\citep{jfm:noack:2003} or center manifold expansions~\citep{GuckenheimerHolmes, jfm:carini:2015}, though in these cases the manifold equation $\bm h$ is usually represented as a Taylor series truncated at second-order.
This is sufficient near bifurcations, but the more general form enabled by sparse regression allows improved resolution of the manifold structure.

%%%%%%%%%%%%%%%%%%%%%%%%%%%%%%
%%%%%     CONCLUSION     %%%%%
%%%%%%%%%%%%%%%%%%%%%%%%%%%%%%
\section{Results}
\label{sec: results}

Despite the near-perfect kinematic resolution of the flow field in the POD basis, there is no level of truncation, up to at least $r=64$, that leads to a Galerkin system that is both stable and reproduces the quasiperiodic dynamics of the flow, as illustrated in Figure~\ref{fig: rom -- pod-galerkin energy evolution}.
In this section we show that polynomial nonlinear correlations can be used to construct a 4-dimensional model that captures the major structure of the post-transient flow.
Based on these results, we argue that accurate kinematic resolution of the advection-dominated flow in the modal basis creates spurious dynamic variables and fragility in the Galerkin systems, as for the linear advection example in Section~\ref{sec: example -- advection}.

\subsection{Nonlinear correlation analysis}
\label{sec: results -- correlations}
As discussed in the DMD analysis of Section~\ref{sec: modes -- dmd}, several of the modal coefficients are nearly pure harmonics of one of the two dominant frequencies, as a result of the the space-time decomposition of traveling wave-type structure in the physical field.
This signature is clear in the phase portraits in Figures~\ref{fig: modes -- harmonics} and~\ref{fig: correlations -- lissajous}, where some coefficient pairs form Lissajous orbits, which are characteristic of harmonic oscillators with frequencies at an integer ratio.
This is reflected in the relatively large scores of the RDC metric of dependence between harmonic mode pairs (\ref{fig: correlations -- lissajous}, upper triangular).
This measure also confirms that the shear layer dynamics and inner cavity motions are nearly independent.

However, the modal analysis in Section~\ref{sec: modes} and the power spectrum in Figure~\ref{fig: plant description -- fourier spectrum} both indicate that the flow cannot be described by purely independent oscillation.
Instead, the flow behaves more like the model in equation~\eqref{eq: correlations -- toy}, where a linear-quadratic system is driven by self-sustaining oscillators at incommensurate frequencies, with higher-order modes connected via cascading nonlinear interactions.
If this is indeed the case, the resulting triadic structure would lead to energy content at frequencies that are not pure harmonics.
In other words, coefficients that are not significantly correlated with the driving oscillators according to the RDC may instead have multivariate nonlinear correlation, or frequency crosstalk.

Based on this intuition, we apply the sparse manifold regression approach described in Section~\ref{sec: correlations -- sparse}.
By applying FROLS with a residual tolerance of $0.1$ we find that DMD coefficients up to $\alpha_{52}$ can be reconstructed with at least 90\% accuracy in a library of polynomials in $\hat{\bm \alpha}$ up to seventh order.
Of these, 24 coefficients can be approximated with residual $10^{-2}-10^{-3}$ with only one polynomial function of the active variables, indicating that the mode approximately is a product of a single triadic interaction.
None of the coefficients require more than five terms (out of a library of 330).

\begin{figure}
\centering
\includegraphics[width=0.9\textwidth]{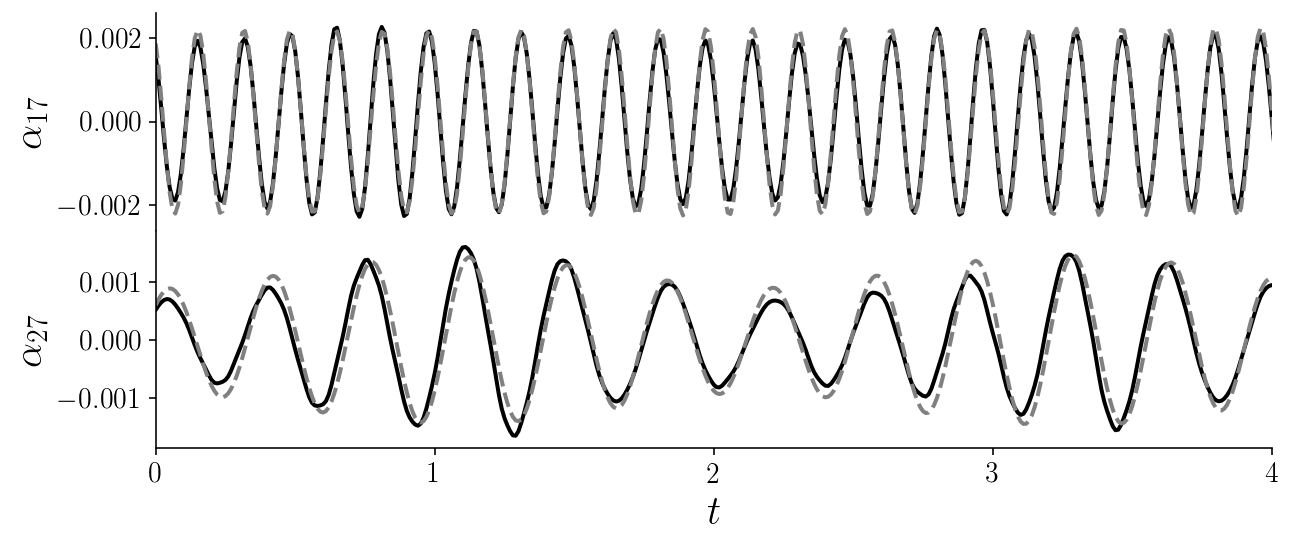}
\caption{
Example coefficient reconstructions $\bm \alpha \approx \bm h (\hat{\bm \alpha})$ based on the leading DMD coefficients (\textbf{---}).
The sparse polynomial approximation ({\color{gray} \textbf{- -}}) for higher-order modes with pure frequency content (e.g. $\alpha_{17} \approx h_{17}(\hat{\bm \alpha})$) tends to be more accurate than for modes with mixed content ($\alpha_{27}$). 
}
\label{fig: results -- manifold-coeff-reconstruction}
\end{figure}

This analysis also reveals that the higher-order coefficients have three distinct relationships to the active degrees of freedom, as illustrated in Figure~\ref{fig: results -- manifold-coeff-reconstruction} and Table~\ref{tab: results -- correlations}:
\begin{enumerate}
    \item Pure harmonics
    \begin{equation}
    \label{eq: results -- frequency-harmonic}
        \alpha_j \propto \alpha_k^{d} {\alpha_k^*}^{d^\prime}, \hspace{1cm} k \in (1, 5).
    \end{equation}
    In this case the dominant frequency will be $\omega_j \sim (d - d')\omega_k$. These coefficients have a high RDC score and have Lissajous-type phase portraits.
    \item Nonlinear crosstalk
    \begin{equation}
    \label{eq: results -- frequency-crosstalk}
        \alpha_j \propto \alpha_1^{c} {\alpha_1^*}^{c^\prime} \alpha_5^{d} {\alpha_5^*}^{d^\prime} 
    \end{equation}
    with dominant frequency $\omega_j \sim (c - c')\omega_s + (d - d')\omega_c$.
    These coefficients have multivariate nonlinear correlation with the active degrees of freedom, so may not have high RDC score.  Two-dimensional phase portraits will also not appear meaningful.
    Still, the coefficients have energy content at a single frequency.
    \item Mixed frequency content: these coefficients cannot be expressed as a single polynomial term in $\hat{\bm \alpha}$, but require 2-5 terms for a reasonably accurate approximation. The coefficients may still be an algebraic function of the active variables (i.e. a sum of terms like~\eqref{eq: results -- frequency-harmonic} and~\eqref{eq: results -- frequency-crosstalk}), but will have energy content at various frequencies.
\end{enumerate}

\begin{table}
\centering
  \begin{tabular}{ c  c  c  c  c }
  Coefficient & Active terms & Triadic frequencies & DMD eigenvalue & RDC vs. $\alpha_1$ \\
  $\alpha_3$ & ${\alpha_1^*}^2 |\alpha_1|^2$ & $ 2\omega_s$ & $ - 0.000 + 23.4 i$ & 0.86 \\
  $\alpha_{17}$ & $\alpha_1^3 \alpha_5$ & $3\omega_s + \omega_c$ &
  $ -0.007 + 37.9 i$ & 0.08 \\
  $\alpha_{27}$ & $\alpha_1 \alpha_5^2$, $\alpha_1 \alpha_5^3$ & $\omega_s + 2 \omega_c $, $ \omega_s + 3 \omega_c $ & 
  $-0.020 + 17.5 i$ & 0.26 \\
  %$\alpha_{51}$ & $\alpha_1^2 |\alpha_1|^2 {\alpha_5^*}^3$, $\alpha_1 |\alpha_1|^2 \alpha_5^2 |\alpha_5|^2 $, $\alpha_1 \alpha_5^3$ & $2 \omega_s - 3 \omega_c $, $ \omega_s + 2 \omega_c $, $ \omega_s + 3 \omega_c $& $ -0.218 + 16.8i $
  \end{tabular}
  \caption{
  Representative nonlinear correlations identified by sparse regression, including pure harmonic ($\alpha_3$), nonlinear crosstalk $(\alpha_{17})$, and mixed frequency content $(\alpha_{27})$.
  Polynomial combinations give rise to oscillations at frequencies in terms of the shear layer $\omega_s \approx 11.7$ and inner cavity $\omega_c \approx 2.7$.
  For modes with nearly pure frequency content (e.g. $\alpha_3, \alpha_{17}$), the resulting frequencies are close to those predicted by the DMD analysis.
  The randomized dependence coefficient (RDC) between the coefficient and $\alpha_1$ is strongest for pure harmonics ($\alpha_3$), even though mixed-frequency modes can be accurately approximated with a simple polynomial function. See figure~\ref{fig: correlations -- lissajous} for pairwise RDC values for the leading 24 DMD coefficients.
  }
  \label{tab: results -- correlations}
\end{table}

As shown by Figure~\ref{fig: results -- manifold-coeff-reconstruction}, the polynomial approximations tend to be more accurate for coefficients with pure frequency content, although they do capture the dominant trends for coefficients with mixed content.
These sparse polynomial representations of higher-order coefficients determine the manifold equation $\bm \alpha = \bm h(\hat{\bm \alpha})$ based on equation~\eqref{eq: correlations -- sindy}.

\subsection{Manifold Galerkin model}
\label{sec: results -- galerkin}
This manifold restriction leads to effective higher-order nonlinearity in the reduced dynamics for $\hat{\bm \alpha}$, given by equation~\eqref{eq: correlations -- manifold-galerkin}.
In particular, since we include up to $7^\mathrm{th}$-order polynomials in the manifold equation, the effective dynamics based on the quadratic Galerkin model involves $14^\mathrm{th}$-order terms.
Fortunately, provided $\mathsfbi{H}$ is sufficiently sparse, the overall cost of evaluating the reduced-order model still only scales with $r^3$ (from $\mathcal{O}(r)$ evaluations of the quadratic term).
The advantage of this additional nonlinearity is that the system is now constrained to the manifold determined by $\bm h(\hat{\bm \alpha})$.
This mitigates both the issue of spurious degrees of freedom in the Galerkin representation of hyperbolic dynamics and the effect of truncating the dissipative scales of the energy cascade.

\begin{figure}
\centering
\includegraphics[width=0.9\textwidth]{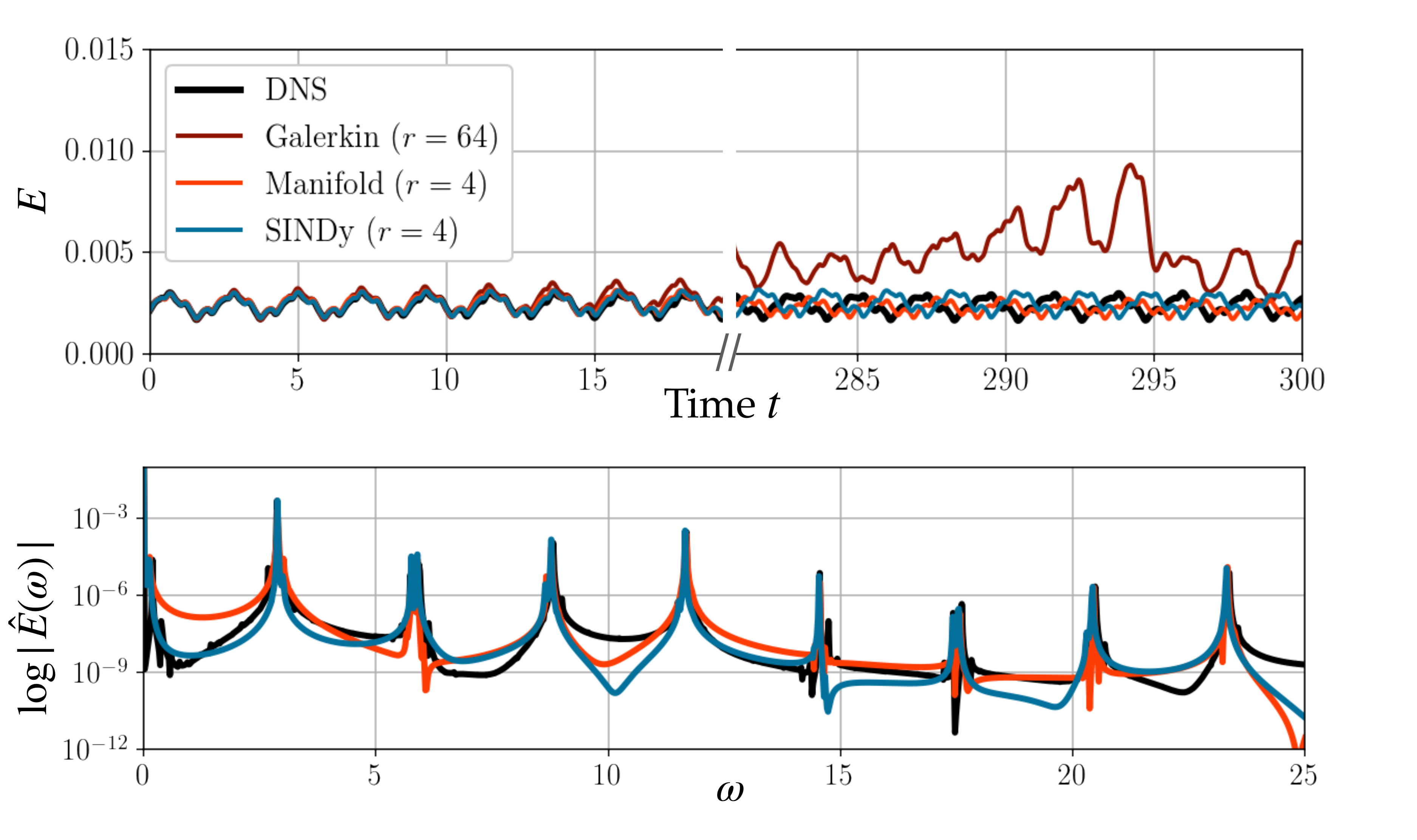}
\caption{
Evolution of the fluctuation kinetic energy for the reduced-order models compared to DNS.
By accounting for nonlinear correlations, both the manifold Galerkin and SINDy models remain at the correct energy level at long times, despite having many fewer degrees of freedom than the standard Galerkin model (top).
Similarly, both models resolve the nonlinear interactions leading to the discrete peaks in the power spectrum (bottom).
}
\label{fig: results -- manifold-energy}
\end{figure}

Simulation results for the manifold Galerkin model are shown in Figure~\ref{fig: results -- manifold-energy} along with the SINDy model discussed in Section~\ref{sec: results -- sindy}.
Whereas the standard Galerkin model eventually overestimates the fluctuation energy and becomes aperiodic, the manifold restriction applied to the same operators continues at the correct energy level and with approximately discrete peaks in the frequency spectrum at the correct locations.
Of course, there is some phase drift for all models at long times, but the manifold reduction prevents the higher-order coefficients from losing coherence with the dominant oscillations and causing the amplitude drift as in the standard Galerkin model.

\subsection{SINDy model}
\label{sec: results -- sindy}

The manifold restriction applied to the Galerkin model results in a significant reduction in dimensionality and improvement in stability and accuracy.
However, the full evolution equation~\eqref{eq: correlations -- manifold-galerkin} is dense with $\mathcal{O}(r^3)$ entries, where $r$ is the size of the POD/DMD subspace, not the number of active degrees of freedom.
This is still a significant improvement over both the DNS and the standard Galerkin model, but the physical picture of coupled nonlinear oscillators giving rise to the quasiperiodic dynamics suggests that a simpler model may capture the dominant features of the flow.

This desire for minimalistic models has been the motivation for several recent applications of SINDy and related system identification techniques to model reduction.
However, when these methods are applied to modal coefficients, they also face the fundamental representation issue challenging Galerkin models.
That is, models with full kinematic resolution will include spurious dynamical degrees of freedom.
The issue is exacerbated in data-driven methods, since both the dimension and the conditioning of the library matrices tend to scale poorly with dimensionality.

For example, it is well known that the flow past a cylinder at Reynolds number 100 can be accurately described by a Stuart-Landau equation with two degrees of freedom.
However, fully reconstructing the post-transient vortex street requires on the order of ten POD modes, all of which are harmonics of the leading pair.
\citet{Loiseau2018} addressed this by identifying the Stuart-Landau equation with SINDy along with a similar sparse regression approach to~\eqref{eq: correlations -- sindy} to algebraically reconstruct the harmonics.

Here we take a similar approach and assume that we do not need the full order-$r$ dynamics and manifold equation to describe the dynamics of the active degrees of freedom $\hat{\bm \alpha}$.
In particular, we anticipate that the minimal description will take the form of coupled Stuart-Landau equations.
As described in Section~\ref{sec: rom -- sindy}, we construct a library of candidate polynomials including up to cubic terms in $\alpha_1, \alpha_1^*, \alpha_5,$ and $\alpha_5^*$.

We identify symbolic equations for $\alpha_1$ and $\alpha_5$ dynamics with the FROLS algorithm, for DMD coefficients $\alpha_2 = \alpha_1^*$ and $\alpha_6 = \alpha_5^*$.
FROLS is an iterative, forward greedy algorithm that requires a stopping condition.
Often a residual-error criterion is used, as in Section~\ref{sec: results -- correlations}, but in this case the DMD modes are so close to pure linear oscillation that a single linear term leaves a residual $\sim 10^{-6}$.
Instead, we stop the iteration at the second term in each equation, retaining a stabilizing cubic term.
The resulting model takes the form
\begin{subequations}
\label{eq: results -- sindy-model}
\begin{align}
    \dot{\alpha}_1 &= \lambda_1 \alpha_1 - \mu_1 \alpha_1|\alpha_1|^2 \\
    \dot{\alpha}_5 &= \lambda_5 \alpha_5 - \mu_5 \alpha_5|\alpha_5|^2,
\end{align}
\end{subequations}
where all $\lambda$ and $\mu$ coefficients are complex.

The system in equation~\eqref{eq: results -- sindy-model} is a pair of independent nonlinear Stuart-Landau oscillators.
Recalling the toy model in Section~\ref{sec: correlations -- model}, independent oscillators that drive a cascade of triadic interactions can lead to quasiperiodic dynamics, even without direct dynamical coupling between the oscillators.
In contrast to the manifold Galerkin model, it is not necessary to reconstruct the full vector of coefficients to solve this minimal system.
Of course, the full vector of coefficients can still be reconstructed after simulating~\eqref{eq: results -- sindy-model} via the manifold function $\bm h$.

\begin{figure}
\centering
\includegraphics[width=0.99\textwidth]{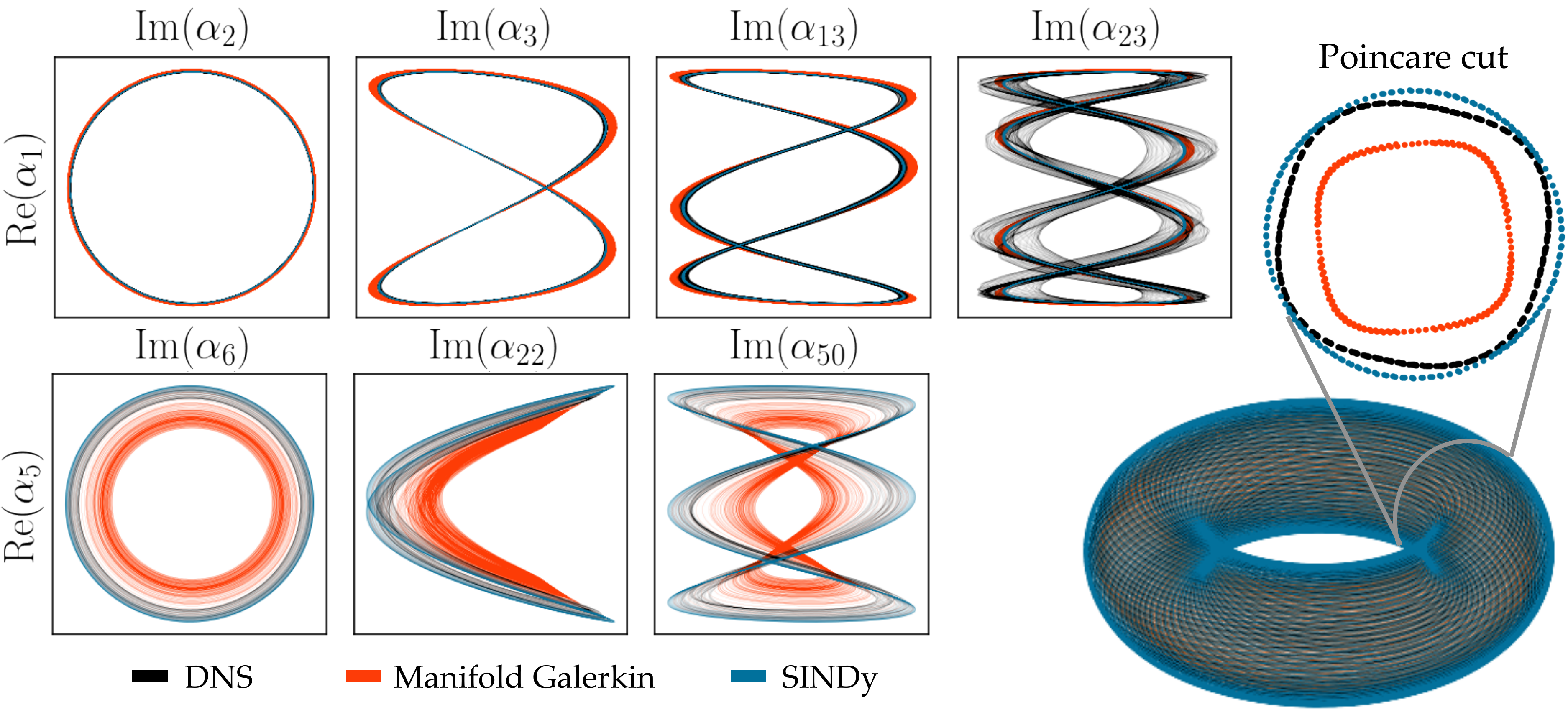}
\caption{
Lissajous figures for the POD coefficients reconstructed from the manifold Galerkin and SINDy reduced-order models (left).
Both models accurately capture the shear layer instability and its harmonics (e.g. $\alpha_1, \alpha_3, \alpha_{13}, \alpha_{23}$), though the manifold Galerkin model tends to underestimate the amplitude of the inner cavity motions (e.g. $\alpha_5, \alpha_{22}, \alpha_{50}$).
A Poincare section of the toroidal attractor confirms this discrepancy, but shows clearly that both models are quasiperiodic and remain on the approximate attractor.
}
\label{fig: results -- rom-lissajous}
\end{figure}

Figure~\ref{fig: results -- manifold-energy} compares the fluctuation kinetic energy based on reconstructions from the SINDy model to the standard and manifold Galerkin models.
As for manifold Galerkin, the SINDy model remains at the correct energy level at long times and reproduces the characteristic structure of the power spectrum.
A slightly more sensitive evaluation is given in Figure~\ref{fig: results -- rom-lissajous}, which compares Lissajous figures of reconstructed near-harmonic POD modes.
Both models accurately capture the harmonics associated with the shear layer instability, but the Galerkin system somewhat underestimates the amplitude of the inner cavity mode and its harmonics.

Although the reduced state space of the pair of complex coefficients is four-dimensional, the quasiperiodic oscillatory nature of the dynamics also offers a convenient symmetry reduction for the purposes of visualization.
With the amplitude-phase representation $\alpha_k = R_k e^{i \phi_k} $, we can approximate the toroidal attractor in the three-dimensional space by representing $\alpha_5$ as an expansion about the point in the complex plane defined by $\alpha_1$:
\begin{subequations}
\begin{align}
    x &= (R_1 + R_2 \cos \phi_2) \cos \phi_1  \\
    y &= (R_1 + R_2 \cos \phi_2) \sin \phi_1  \\
    z &= R_2 \sin \phi_2.
\end{align}
\end{subequations}
Finally, the models can be compared in detail with a Poincare section of this torus about any convenient plane (we choose $x=0$); both the three-dimensional phase portrait and Poicare section are also shown in Figure~\ref{fig: results -- rom-lissajous}.

As a result of the slight underestimation of energy in the inner cavity motions, the Poincare section for the manifold Galerkin model shows a somewhat smaller attractor than the DNS and SINDy.
Conversely, the highly simplified structure of the SINDy model leads to a circular section, while the Galerkin system captures the rounded-square shape of the true section.
This is likely a consequence of the high-order effective nonlinearity in the Galerkin system, which allows it to resolve more complex attractor shapes.
Nevertheless, the SINDy system does give an accurate estimate of the typical amplitude in the slice and preserves the coherence of the harmonic modes for both the shear layer and inner cavity oscillations.

% What's the upshot?  Preview discussion.
In both the manifold Galerkin and SINDy models, the nonlinear correlations play a critical role in the accuracy and stability of the reduced-order dynamics.
The space-time decomposition of an advection-dominated flow introduces a significant number of modes that are necessary to reconstruct the field, but do not correspond to independent degrees of freedom in the dynamics.
Nonlinear correlations analysis provides a straightforward, principled approach to restricting the Galerkin dynamics to the set of active degrees of freedom, as well as convenient coordinates for system identification.

%%%%%%%%%%%%%%%%%%%%%%%%%%%%%%
%%%%%     CONCLUSION     %%%%%
%%%%%%%%%%%%%%%%%%%%%%%%%%%%%%
\section{Discussion}
\label{sec: discussion}
It has been widely recognized for some time that Galerkin-type models of advection-dominated flows are prone to fragility and instability.
The majority of work addressing this issue has focused on truncation of the energy cascade, leading to closures in the vein of subgrid-scale large eddy simulation models~\citep{jfm:rempfer:1994, cmame:wang:2012, Cordier2013, jfm:osth:2014, pre:san:2018, Pan2018siam}.
However, recent work including~\citet{jcp:carlberg:2017, Grimberg2020jcp, Lee2020jcp} has begun questioning the fundamental suitability of Galerkin projection for hyperbolic problems, pointing out for instance that any notion of optimality associated with the Galerkin system is lost upon time discretization.
This perspective is perhaps supported by the observation that Galerkin-type reduced-order models often do not significantly improve with increasing rank, as one might expect if the primary issue was under-resolved dissipation.

In this work we have used a nonlinear correlations analysis of a quasiperiodic shear-driven cavity flow to argue that decoherence resulting from the linear modal representation of advecting structures also deserves consideration.
This space-time decomposition introduces one temporal coefficient per spatial mode; in many cases this may result in many more coefficients than there are degrees of freedom in the post-transient flow.
Galerkin models treat each coefficient as an independent degree of freedom; small errors in the system of differential equations can lead to catastrophic decoherence and instability.
Instead, we show that exploiting statistical structure and algebraic dependence in the temporal coefficients enables the reduction of the dynamical system to the true rank while preserving the kinematic resolution of the modal basis.

The cavity flow is dominated by two key modal structures: a high-frequency shear layer instability and low-frequency inner cavity oscillation.
The natural dynamics of the flow are quasiperiodic, as can be seen from the characteristic power spectrum in Figure~\ref{fig: plant description -- fourier spectrum}.
Both the shear layer and inner cavity features can be identified by a stability analysis of the time-averaged mean flow (see Appendix~\ref{appendix: linear stability}).
However, linear modal representations (POD or DMD) approximate the traveling wave structures in the nonlinear flow with not only the fundamental stability modes, but also higher harmonics and nonlinear crosstalk modes, each of which can be associated with one of the approximately discrete peaks in the power spectrum.

The physical coherence (non-dispersion) of the fundamental flow features appears as nonlinear correlation between temporal coefficients associated with harmonics and crosstalk.
This can also be conceptualized as triadic interactions in the frequency domain.
After using a randomized dependence coefficient (RDC) analysis to identify the dynamically active modes, we use sparse polynomial regression to uncover simple algebraic relationships that account for $\sim 99.5\%$ of the fluctuation kinetic energy.

We give examples of two ways in which these relationships can be used to improve reduced-order models.
First, they act as a simple manifold equation constraining the Galerkin dynamics to the post-transient attractor; this may be viewed as a data-driven generalization of analytic invariant manifold reductions (e.g.~\citet{jfm:noack:2003}), which rely on scale-separation arguments.
Alternatively, the driving coefficients offer a convenient basis for nonlinear system identification; we use the SINDy framework~\citep{pnas:brunton:2016} to identify a simple model of the flow as a pair of independent Stuart-Landau oscillators.
The full flow field may then be reconstructed with the manifold equation.

The manifold Galerkin system connects the reduced-order system the governing equations and may allow for more natural parametric variation, but requires accurate estimates of the gradients of the POD modes and/or intrusive access to the full-order solver.
On the other hand, the SINDy model is compact, non-intrusive, and more amenable to analytical treatment, though it cannot be directly connected to the underlying physical equations and it is more difficult to capture parametric variation.
In general, the most appropriate choice is likely to be application-dependent.

Regardless of the chosen model reduction technique, we conclude that exploiting nonlinear structure in the modal coefficients is a natural and efficient approach to improving the stability and accuracy of low-order models.
In a broader sense, our approach to this analysis (with the RDC and sparse polynomial regression) can be seen as simple, interpretable manifold learning.
This is sufficient for quasiperiodic dynamics, since the form of the nonlinear dependence can be readily deduced by reasoning about the triadic interactions.
In the language of Koopman theory, the flow has a discrete or point spectrum~\citep{arfm:mezic:2013,Arbabi2017prf}; more sophisticated analysis would be necessary to extend these results to chaotic or turbulent systems with continuous spectra.

However, in these cases any invertible manifold learning method could be used to the same end.
This might include deep learning techniques such as autoencoder networks~\citep{ieee:bengio:2013}, an unsupervised method that learns a compressed representation of high-dimensional data.
Autoencoders have recently been explored for black-box forecasting~\citep{prsa:vlachas:2018}, system identification~\citep{Champion2019pnas}, and model reduction~\citep{Lee2020jcp}.
Regardless of the method, nonlinear embedding recognizes the intrinsic dimensionality of the dynamics as distinct from that of the linear subspace required to reconstruct the flow field.

There are also many opportunities for further work in reduced-order model development.
For instance, accounting for nonlinear correlations does not address the issue of truncating the energy cascade.
This does not pose a problem for the present laminar, two-dimensional flow, but severe dimensionality reduction of any multiscale or turbulent dynamics will necessarily act as a spatiotemporal filter.
In other words, the dissipative scales will generally not be correlated in any way with the large-scale dynamics and a closure strategy is likely to be necessary in order to accurately capture the dissipation rate.
This is typically less of an issue in the system identification framework, since the natural dynamics are estimated at once, including any effective closure models within the span of the candidate functions.
However, for more complex flows it may be necessary to employ a more sophisticated optimization~\citep{Champion2019ieee}, physics-based constraints~\citep{jfm:loiseau:2018a}, or enforcement of long-term stability~\citep{jfm:schlegel:2015, Kaptanoglu2021}.
Despite prospective challenges in scaling this approach to chaotic and turbulent flows, we expect that there are significant stability and robustness benefits to be realized by exploiting nonlinear correlations in reduced-order models of coherent structures in advection-dominated flows.

\section*{Acknowledgements}
JLC acknowledges funding support from the Department of Defense (DoD) through the National Defense Science \& Engineering Graduate (NDSEG) Fellowship Program.  
SLB acknowledges funding support from the Army Research Office (ARO W911NF-19-1-0045; program manager Dr. Matthew Munson).
The authors are grateful for discussions with Nathan Kutz and Georgios Rigas.

\clearpage
%%%%%%%%%%%%%%%%%%%%%%%%%%%%
%%%%%     APPENDIX     %%%%%
%%%%%%%%%%%%%%%%%%%%%%%%%%%%
\appendix
\section{Linear stability analysis}
\label{appendix: linear stability}

   \begin{figure}
     \centering
     \includegraphics[width=\textwidth]{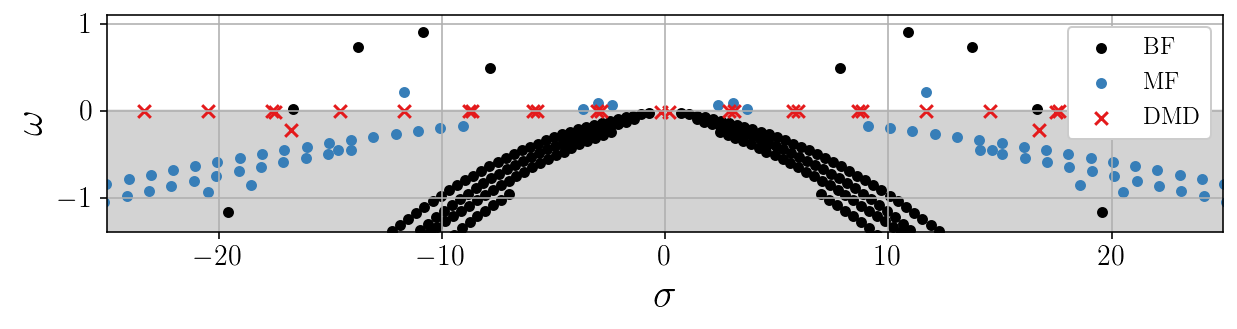}
    %  \subfigure[]{  \hspace{-.35in}\includegraphics[width=.84\textwidth]{Eigenspectrum_re7500_baseflow}}\\
    %  \subfigure[]{  \hspace{-.5in}\includegraphics[width=.9\textwidth]{mean_flow_eigenspectrum}}
     \caption{Eigenspectrum of the Navier-Stokes operator at $Re=7500$ estimated in three ways: linearized in the vicinity of the base flow (BF, black circles), mean flow (MF, blue circles), and from dynamic mode decomposition (DMD, red crosses). For the linear stability analyses, only eigenvalues for which a $10^{-8}$ convergence has been achieved are plotted.  These eigenspectra have been computed using a time-stepper Arnoldi algorithm with a sampling period $\Delta T=0.1$ and a Krylov subspace dimension of 1024 and 512 for the base and mean flows, respectively.
     See Section~\ref{sec: modes -- dmd} for details on the DMD analysis.}
     \label{fig: appendix -- eigenspectra}
   \end{figure}

  A key assumption in the present work is that analysis of the Navier-Stokes operator linearized about the unstable steady state provides very limited insights into the post-transient nonlinear dynamics.
  This precludes a number of powerful analytic tools such as center manifold analysis~\citep{jfm:carini:2015} and multiple scale expansions~\citep{jfm:sipp:2007}, leaving semi-empirical or fully data-driven methods.
  In order to support this assumption, this appendix briefly summarizes the results of such a linear stability analysis.
  
  %Although not demonstrated, it has been assumed throughout the present work that investigating the properties of the Navier-Stokes operator linearized in the vicinity of the true base flow provides very limited insights into the frequencies observed in direct numerical simulation.
  %As to support our claim, this appendix briefly summarizes the results of such a linear stability analysis.
  
  Denoting the full flow field by $\bm q(\bm x) = \begin{bmatrix} \bm u(\bm x) & p(\bm x) \end{bmatrix}^T$, the base flow $\bm q_b$ is the solution to the steady-state Navier-Stokes equations
%   \begin{subequations}
%     \begin{align}
%       \displaystyle \nabla \cdot \left( \bm{u}_b \otimes \bm{u}_b \right)  &= - \nabla p_b + \frac{1}{\Rey} \nabla^2 \bm{u}_b,  \\
%       \nabla \cdot \bm{u}_b  &= 0.
%     \end{align}
%     \label{eq: ns-ss}
%   \end{subequations}
    \begin{equation}
      \nabla \cdot \left( \bm{u}_b \otimes \bm{u}_b \right) = - \nabla p_b + \frac{1}{\Rey} \nabla^2 \bm{u}_b,  \qquad
      \nabla \cdot \bm{u}_b  = 0.
    \label{eq: ns-ss}
  \end{equation}
  Since this solution is linearly unstable, we approximate it with the selective frequency damping algorithm~\citep{pof:akervik:2006}.
  The linearized equations for the evolution of an infinitesimal perturbation $\bm q^\prime(\bm x, t)$ are
   %
%   \begin{subequations}
%     \begin{align}
%      \displaystyle \frac{\partial \bm{u}^{\prime}}{\partial t}  &= - \left( \bm{u}^{\prime} \cdot \nabla \right) \bm{u}_b - \left( \bm{u}_b \cdot \nabla \right) \bm{u}^{\prime} - \nabla p^{\prime} + \frac{1}{\Rey} \nabla^2 \bm{u}^{\prime} \\
%       \nabla \cdot \bm{u}^{\prime} &= 0.
%     \end{align}
%     \label{eq: ns-lin}
%   \end{subequations}

  \begin{equation}
     \displaystyle \frac{\partial \bm{u}^{\prime}}{\partial t}  = - \left( \bm{u}^{\prime} \cdot \nabla \right) \bm{u}_b - \left( \bm{u}_b \cdot \nabla \right) \bm{u}^{\prime} - \nabla p^{\prime} + \frac{1}{\Rey} \nabla^2 \bm{u}^{\prime}, \qquad
      \nabla \cdot \bm{u}^{\prime} = 0.
    \label{eq: ns-lin}
  \end{equation}
   Introducing a normal mode ansatz
   $$
     \bm{q}^{\prime}(\bm{x}, t) = \hat{\bm{q}}(\bm{x}) e^{(\sigma + i\omega)t}
   $$
   where $\bm{q}^\prime = \begin{bmatrix} \bm{u}^{\prime} & p^{\prime} \end{bmatrix}^T$, the linearized Navier-Stokes equations can be recast as a generalized eigenvalue problem
   $$
   \left( \sigma + i \omega \right) \mathsfbi{B} \hat{\bm{q}} = \mathsfbi{L} \hat{\bm{q}},
   $$
   where $\mathsfbi{L}$ is the Jacobian matrix of the Navier-Stokes equations and $\mathsfbi{B}$ is the singular mass matrix.
   The leading eigenpairs of the pencil $(\mathsfbi{A}, \mathsfbi{B})$ are then computed using an in-house Krylov-Schur time-stepping algorithm~\citep{jcp:edwards:1994, siam:stewart:2001} implemented in the spectral element solver Nek5000 \citep{nek5000_site}.
   For more details, see \citet{jcp:edwards:1994, siam:stewart:2001, aiaa:bagheri:2009, amr:sipp:2010} or the recent review chapter \citet{chapter:loiseau:2019}.
  
  Figure \ref{fig: appendix -- eigenspectra} depicts the eigenspectra of the operator linearized about base and  flows.
  Four complex conjugate pairs of eigenvalues lie within the upper-half complex plane, indicating the base flow is strongly unstable.
  The most unstable eigenvalue is
  $
      \sigma \pm i \omega = 0.90 \pm 10.86 i,
  $
  where $\sigma$ and $\omega$ are the growth rate and circular frequency of the instability mode, respectively.
  This frequency differs by only 5\% to 10\% from the dominant peak of the DNS (Figure~\ref{fig: plant description -- fourier spectrum}) and that given by the DMD analysis; leading eigenvalues are compared in Table~\ref{tab: appendix -- eigvals}.
  The associated eigenfunction (not shown) also closely resembles the leading DMD mode $\bm \phi_1$ shown in Figures~\ref{fig: modes -- harmonics} and~\ref{fig: modes -- dmd-spectrum}.
  
\begin{table}
\centering
  \begin{tabular}{ c  c  c  c c }
  Base flow & $0.90 \pm 10.86 i$ & $0.73 \pm 13.75 i$ & $0.49 \pm 7.85 i$ &  $0.02 \pm 16.65 i$ \\
  Mean flow  & $0.22 \pm 11.68 i$ & $0.09 \pm 3.03 i$ & $0.06 \pm 2.38 i$ &  $0.03 \pm 14.68 i$\\
  DMD  & $0.00 \pm 11.68 i$ & $0.00 \pm 23.36 i$ & $0.00 \pm 2.87 i$ &  $0.02 \pm 3.68 i$ \\
  \end{tabular}
  \caption{
  Eigenvalues $\sigma + i \omega$ of the least stable modes in a stability analysis of the base and mean flows, along with DMD eigenvalues for the most energetic modes.
  }
  \label{tab: appendix -- eigvals}
\end{table}
  
Although the stability analysis of the base flow provides some insight about the physical origin of the high-frequency shear layer oscillation, there are two main reasons it is insufficient to describe the nonlinear flow.
%First of all, despite the existence of three additional unstable pairs of complex-conjugate eigenvalues, no trace of the associated oscillatory dynamics has been observed in direct numerical simulation.
% More importantly, for the Reynolds number considered, linear stability analysis of the base flow is completely unable to predict the existence of the low-frequency dynamics occurring within the cavity.
First, there is no trace of the three additional unstable modes predicted by the stability analysis in the direct numerical simulation.
Second, at the Reynolds number considered in this work, linear stability analysis of the base flow is unable to predict the low-frequency inner-cavity oscillation.
The higher harmonics are also missing from the stability analysis, though this is to be expected of a linear analysis.
To the authors' knowledge, there has not yet been a detailed explanation of the process by which the additional unstable modes are superseded by the low-frequency dynamics.
For more details about the shear layer instability and its saturation process at lower Reynolds number, see \citet{jfm:meliga:2017}.

%Despite both of these observations being captured by performing a linear stability analysis in the vicinity of the mean flow, no clear explanation can be given yet as to why we do not observe the frequency signature of the additional unstable eigenvalues, nor about the precise origin of the low-frequency unsteadiness observed within the cavity. Our attention is currently devoted to these two problems.
%

Although it is not a steady-state solution of the Navier-Stokes equations, several studies have shown that linearizing about the mean flow $\bar{\bm{u}}(\bm{x})$ nonetheless provides valuable insights into the dynamics of coherent structures existing in the nonlinear flow \citep{jfm:malkus:1956, epl:barkley:2006, prl:mantic:2014, jfm:beneddine:2016, jfm:meliga:2017}.
The analysis is the same as above, replacing the base flow $\bm q_b(\bm x)$ with the mean flow $\bar{\bm q}(\bm x)$.
%The resulting spectrum is also shown in Figure~\ref{fig: appendix -- eigenspectra}, with leading eigenvalues also given in Table~\ref{tab: appendix -- eigvals}.
In contrast to the base flow analysis, stability analysis of the mean flow does predict both the shear layer instability and inner-cavity oscillations.
%The harmonics, which are generated by nonlinear interactions, are still not resolved by the linear stability analysis.

This improvement of the stability analysis about the mean flow accounts for its increasing popularity in both modal~\citep{jfm:sipp:2007, jfm:beneddine:2016} and nonmodal~\citep{McKeon2010jfm} analysis.
It is also appealing experimentally, since the mean flow can be estimated practically for statistically stationary flows, while unstable steady states are difficult to produce.
However, from a numerical perspective standard mean flow analysis is not predictive in the sense that fully-converged statistics are necessary to compute the mean flow prior to the stability analysis.  
%In that scenario POD or DMD analysis may be more revealing and does not require access to the linearized operator.
Predictive mean flow analysis is the subject of ongoing work, for example with eddy viscosity-based Reynolds-averaged Navier-Stokes mean flow estimates~\citep{Pickering2020} or self-consistent modeling~\citep{prl:mantic:2014, jfm:meliga:2017}.

In this case, the mean flow stability analysis supports the picture suggested by the nonlinear correlations analysis; the four active degrees of freedom are related to the two mode pairs corresponding to the shear layer instability and inner cavity oscillation.
In the nonlinear DNS, interactions between these modes generate harmonics and frequency crosstalk, although this structure is fully dependent on the active degrees of freedom.
Linear stability analysis assumes the perturbations have negligible energy and so it cannot resolve the nonlinear interactions responsible for this behavior.

\newpage
\bibliography{bibliography}% Produces the bibliography via BibTeX.

\begin{thebibliography}{102}
\expandafter\ifx\csname natexlab\endcsname\relax\def\natexlab#1{#1}\fi
\def\au#1{#1} \def\ed#1{#1} \def\yr#1{#1}\def\at#1{#1}\def\jt#1{\textit{#1}}
  \def\bt#1{#1}\def\bvol#1{\textbf{#1}} \def\vol#1{#1} \def\pg#1{#1}
  \def\publ#1{#1}\def\arxiv#1{#1}\def\org#1{#1}\def\st#1{\textit{#1}}

\bibitem[{\AA}kervik {\em et~al.\/}(2006){\AA}kervik, Brandt, Henningson,
  H{\oe}pffner, Marxen \& Schlatter]{pof:akervik:2006}
{\sc \au{{\AA}kervik, E.}, \au{Brandt, L.}, \au{Henningson, D.~S.},
  \au{H{\oe}pffner, J.}, \au{Marxen, O.} \& \au{Schlatter, P.}} \yr{2006}
  \at{Steady solutions of the navier-stokes equations by selective frequency
  damping}.  \jt{Phys. Fluids}  \bvol{18}~(6),  \pg{068102}.

\bibitem[Antoulas(2005)]{Antoulas2005}
{\sc \au{Antoulas, A.~C.}} \yr{2005} {\em Approximation of Large-Scale
  Dynamical Systems\/}.  \publ{SIAM}.

\bibitem[Arbabi \& Mezi\'c(2017)]{Arbabi2017prf}
{\sc \au{Arbabi, H.} \& \au{Mezi\'c, I.}} \yr{2017}  \at{Study of dynamics in
  post-transient flows using {Koopman} mode decomposition}.  \jt{Physical
  Review Fluids}  \bvol{2},  \pg{124402}.

\bibitem[Aubry {\em et~al.\/}(1988)Aubry, Holmes, Lumley \&
  Stone]{jfm:aubry:1988}
{\sc \au{Aubry, N.}, \au{Holmes, P.}, \au{Lumley, J.~L.} \& \au{Stone, E.}}
  \yr{1988}  \at{The dynamics of coherent structures in the wall region of a
  turbulent boundary layer}.  \jt{J. Fluid Mech.}  \bvol{192}~(-1),  \pg{115}.

\bibitem[Bagheri {\em et~al.\/}(2009)Bagheri, {\AA}kervik, Brandt \&
  Henningson]{aiaa:bagheri:2009}
{\sc \au{Bagheri, S.}, \au{{\AA}kervik, E.}, \au{Brandt, L.} \& \au{Henningson,
  D.~S.}} \yr{2009}  \at{Matrix-free methods for the stability and control of
  boundary layers}.  \jt{{AIAA} J.}  \bvol{47}~(5),  \pg{1057--1068}.

\bibitem[Balajewicz {\em et~al.\/}(2013)Balajewicz, Dowell \&
  Noack]{jfm:balajewicz:2013}
{\sc \au{Balajewicz, M.~J.}, \au{Dowell, E.~H.} \& \au{Noack, B.~R.}} \yr{2013}
   \at{{Low-dimensional modelling of high-Reynolds-number shear flows
  incorporating constraints from the Navier{\textendash}Stokes equation}}.
  \jt{J. Fluid Mech.}  \bvol{729},  \pg{285--308}.

\bibitem[Barbagallo {\em et~al.\/}(2009)Barbagallo, Sipp \&
  Schmid]{jfm:barbagallo:2009}
{\sc \au{Barbagallo, A.}, \au{Sipp, D..} \& \au{Schmid, P.~J.}} \yr{2009}
  \at{Closed-loop control of an open cavity flow using reduced-order models}.
  \jt{J. Fluid Mech.}  \bvol{641},  \pg{1}.

\bibitem[Barkley(2006)]{epl:barkley:2006}
{\sc \au{Barkley, D.}} \yr{2006}  \at{Linear analysis of the cylinder wake mean
  flow}.  \jt{Europhys. Lett.}  \bvol{75}~(5),  \pg{750--756}.

\bibitem[Beneddine {\em et~al.\/}(2016)Beneddine, Sipp, Arnault, Dandois \&
  Lesshafft]{jfm:beneddine:2016}
{\sc \au{Beneddine, S.}, \au{Sipp, D.}, \au{Arnault, A.}, \au{Dandois, J.} \&
  \au{Lesshafft, L.}} \yr{2016}  \at{Conditions for validaty of mean flow
  stability analysis}.  \jt{J. Fluid Mech.}  \bvol{798},  \pg{485--504}.

\bibitem[Bengana {\em et~al.\/}(2019)Bengana, Loiseau, Robinet \&
  Tuckerman]{Bengana2019jfm}
{\sc \au{Bengana, Y.}, \au{Loiseau, J.-C.}, \au{Robinet, J.-C.} \&
  \au{Tuckerman, L.~S.}} \yr{2019}  \at{Bifurcation analysis and frequency
  prediction in shear-driven cavity flow}.  \jt{Journal of Fluid Mechanics}
  \bvol{875},  \pg{725--757}.

\bibitem[Bengio {\em et~al.\/}(2013)Bengio, Courville \&
  Vincent]{ieee:bengio:2013}
{\sc \au{Bengio, Y.}, \au{Courville, A.} \& \au{Vincent, P.}} \yr{2013}
  \at{{Representation Learning: A Review and New Perspectives}}.  \jt{IEEE
  Trans. Pattern Anal. Mach. Intell.}  \bvol{35},  \pg{1798--1828}.

\bibitem[Benner {\em et~al.\/}(2015)Benner, Gugercin \&
  Willcox]{Benner2015siam}
{\sc \au{Benner, P.}, \au{Gugercin, S.} \& \au{Willcox, K.}} \yr{2015}  \at{A
  survey of projection-based model reduction methods for parametric dynamical
  systems}.  \jt{SIAM Review}  \bvol{57}~(4),  \pg{483--531}.

\bibitem[Billings(2013)]{book:billings:2013}
{\sc \au{Billings, S.~A.}} \yr{2013} {\em {Nonlinear System Identification:
  NARMAX Methods in the Time, Frequency, and Spatio-Temporal Domains}\/}.
  \publ{Paperbackshop UK Import}.

\bibitem[Brunton {\em et~al.\/}(2021)Brunton, Budi{\v{s}}i{\'c}, Kaiser \&
  Kutz]{Brunton2021koopman}
{\sc \au{Brunton, Steven~L}, \au{Budi{\v{s}}i{\'c}, Marko}, \au{Kaiser, Eurika}
  \& \au{Kutz, J~Nathan}} \yr{2021}  \at{Modern {K}oopman theory for dynamical
  systems}.  \jt{arXiv preprint arXiv:2102.12086} .

\bibitem[Brunton \& Noack(2015)]{amr:brunton:2015}
{\sc \au{Brunton, S.~L.} \& \au{Noack, B.~R.}} \yr{2015}  \at{Closed-loop
  turbulence control: Progress and challenges}.  \jt{Appl. Mech. Rev.}
  \bvol{67}~(5),  \pg{050801}.

\bibitem[Brunton {\em et~al.\/}(2016)Brunton, Proctor \&
  Kutz]{pnas:brunton:2016}
{\sc \au{Brunton, S.~L.}, \au{Proctor, J.~L.} \& \au{Kutz, J.~N.}} \yr{2016}
  \at{Discovering governing equations from data by sparse identification of
  nonlinear dynamical systems}.  \jt{Proc. Natl. Acad. Sci. U.S.A.}
  \bvol{113}~(15),  \pg{3932--3937}.

\bibitem[Carini {\em et~al.\/}(2015)Carini, Auteri \&
  Giannetti]{jfm:carini:2015}
{\sc \au{Carini, M.}, \au{Auteri, F.} \& \au{Giannetti, F.}} \yr{2015}
  \at{Centre-manifold reduction of bifurcating flows}.  \jt{J. Fluid Mech.}
  \bvol{767},  \pg{109--145}.

\bibitem[Carlberg {\em et~al.\/}(2017)Carlberg, Barone \&
  Antil]{jcp:carlberg:2017}
{\sc \au{Carlberg, K.}, \au{Barone, M.} \& \au{Antil, H.}} \yr{2017}
  \at{{Galerkin v. least-squares Petrov{\textendash}Galerkin projection in
  nonlinear model reduction}}.  \jt{J. Comput. Phys.}  \bvol{330},
  \pg{693--734}.

\bibitem[Champion {\em et~al.\/}(2019{\natexlab{{\em a\/}}})Champion, Lusch,
  Kutz \& Brunton]{Champion2019pnas}
{\sc \au{Champion, Kathleen}, \au{Lusch, Bethany}, \au{Kutz, J.~Nathan} \&
  \au{Brunton, Steven~L.}} \yr{2019{\natexlab{{\em a\/}}}}  \at{Data-driven
  discovery of coordinates and governing equations}.  \jt{Proceedings of the
  National Academy of Sciences}  \bvol{116}~(45),  \pg{22445--22451}.

\bibitem[Champion {\em et~al.\/}(2019{\natexlab{{\em b\/}}})Champion, Zheng,
  Aravkin, Brunton \& Kutz]{Champion2019ieee}
{\sc \au{Champion, K.}, \au{Zheng, P.}, \au{Aravkin, A.}, \au{Brunton, S.~L.}
  \& \au{Kutz, J.~N.}} \yr{2019{\natexlab{{\em b\/}}}}  \at{A unified sparse
  optimization framework to learn parsimonious physics-informed models from
  data}.  \jt{IEEE Access}  \bvol{8},  \pg{169259--169271}.

\bibitem[Chien {\em et~al.\/}(1986)Chien, Rising \& Ottino]{jfm:chien:1986}
{\sc \au{Chien, W.-L.}, \au{Rising, H.} \& \au{Ottino, J.~M.}} \yr{1986}
  \at{Laminar mixing and chaotic mixing in several cavity flows}.  \jt{J. Fluid
  Mech.}  \bvol{170}~(-1),  \pg{355}.

\bibitem[Cordier {\em et~al.\/}(2013)Cordier, Noack, Tissot, Lehnasch,
  Delville, Balajewicz, Daviller \& Niven]{Cordier2013}
{\sc \au{Cordier, L.}, \au{Noack, B.~R.}, \au{Tissot, G.}, \au{Lehnasch, G.},
  \au{Delville, J.}, \au{Balajewicz, M.}, \au{Daviller, G.} \& \au{Niven,
  R.~K.}} \yr{2013}  \at{Identification strategies for model-based control}.
  \jt{Experiments in Fluids}  \bvol{54},  \pg{1580}.

\bibitem[Cross \& Hohenberg(1993)]{Cross1993}
{\sc \au{Cross, M.~C.} \& \au{Hohenberg, P.~C.}} \yr{1993}  \at{Pattern
  formation outside of equilibrium}.  \jt{Reviews of modern physics}
  \bvol{65}~(3),  \pg{851}.

\bibitem[Deng {\em et~al.\/}(2020)Deng, Noack, Morzynski \&
  Pastur]{Deng2020jfm}
{\sc \au{Deng, N.}, \au{Noack, B.~R.}, \au{Morzynski, M.} \& \au{Pastur,
  L.~R.}} \yr{2020}  \at{Low-order model for successive bifurcations of the
  fluidic pinball}.  \jt{Journal of Fluid Mechanics}  \bvol{884},  \pg{A37--1}.

\bibitem[Edwards {\em et~al.\/}(1994)Edwards, Tuckerman, Friesner \&
  Sorensen]{jcp:edwards:1994}
{\sc \au{Edwards, W.~S.}, \au{Tuckerman, L.~S.}, \au{Friesner, R.~A.} \&
  \au{Sorensen, D.~C.}} \yr{1994}  \at{Krylov methods for the incompressible
  navier-stokes equations}.  \jt{J. Comput. Phys.}  \bvol{110}~(1),
  \pg{82--102}.

\bibitem[Fischer {\em et~al.\/}(2008)Fischer, Lottes \&
  Kerkemeir]{nek5000_site}
{\sc \au{Fischer, P.F.}, \au{Lottes, J.W.} \& \au{Kerkemeir, S.G.}} \yr{2008}
  {N}ek5000 {W}eb pages. Http://nek5000.mcs.anl.gov.

\bibitem[Glaz {\em et~al.\/}(2017)Glaz, Mezi\'c, Fonoberova \&
  Loire]{Glaz2017jfm}
{\sc \au{Glaz, B.}, \au{Mezi\'c, I.}, \au{Fonoberova, M.} \& \au{Loire, S.}}
  \yr{2017}  \at{Quasi-periodic intermittency in oscillating cylinder flow}.
  \jt{Journal of Fluid Mechanics}  \bvol{828},  \pg{680--707}.

\bibitem[Golubitsky \& Langford(1988)]{Golubitsky1988}
{\sc \au{Golubitsky, M.} \& \au{Langford, W.}} \yr{1988}  \at{Pattern formation
  and bistability in flow between counterrotating cylinders}.  \jt{Physica D}
  \bvol{32},  \pg{362--392}.

\bibitem[Grimberg {\em et~al.\/}(2020)Grimberg, Farhat \&
  Youkilis]{Grimberg2020jcp}
{\sc \au{Grimberg, S.}, \au{Farhat, C.} \& \au{Youkilis, N.}} \yr{2020}  \at{On
  the stability of projection-based model order reduction for
  convection-dominated laminar and turbulent flows}.  \jt{Journal of
  Computational Physics}  \bvol{419},  \pg{109681}.

\bibitem[Guckenheimer \& Holmes(1983)]{GuckenheimerHolmes}
{\sc \au{Guckenheimer, J.} \& \au{Holmes, P.}} \yr{1983} {\em Nonlinear
  oscillations, dynamical systems, and bifurcations of vector fields\/}.
  \publ{Springer}.

\bibitem[Haken(1983)]{Haken1983}
{\sc \au{Haken, H.}} \yr{1983} {\em Synergetics: an Introduction:
  Nonequilibrium Phase Transitions and Self-Organization in Physics, Chemistry
  and Biology\/}.  \publ{Springer New York}.

\bibitem[Holmes {\em et~al.\/}(1996)Holmes, Lumley \&
  Berkooz]{book:holmes:1996}
{\sc \au{Holmes, P.}, \au{Lumley, J.~L.} \& \au{Berkooz, G.}} \yr{1996} {\em
  Turbulence, Coherent Structures, Dynamical Systems and Symmetry\/}.
  \publ{Cambridge University Press}.

\bibitem[Kaptanoglu {\em et~al.\/}(2021)Kaptanoglu, Callaham, Hansen, Aravkin
  \& Brunton]{Kaptanoglu2021}
{\sc \au{Kaptanoglu, A.~A.}, \au{Callaham, J.~L.}, \au{Hansen, C.~J.},
  \au{Aravkin, A.} \& \au{Brunton, S.~L.}} \yr{2021}  \at{Promoting global
  stability in data-driven models of quadratic nonlinear dynamics}.
  \jt{arXiv:2105.01843} .

\bibitem[Kutz {\em et~al.\/}(2016)Kutz, Brunton, Brunton \&
  Proctor]{book:kutz:2016}
{\sc \au{Kutz, J.~N.}, \au{Brunton, S.~L.}, \au{Brunton, B.~W.} \& \au{Proctor,
  J.~L.}} \yr{2016} {\em Dynamic Mode Decomposition: Data-Driven Modeling of
  Complex Systems\/}.  \publ{SIAM-Society for Industrial and Applied
  Mathematics}.

\bibitem[Landau(1944)]{Landau1944}
{\sc \au{Landau, L.~D.}} \yr{1944}  \at{On the problem of turbulence}.
  \jt{Doklady Akademiia Nauk SSSR}  \bvol{44},  \pg{339--342}.

\bibitem[Leclercq {\em et~al.\/}(2019)Leclercq, Demourant, Poussot-Vassal \&
  Sipp]{jfm:leclerc:2019}
{\sc \au{Leclercq, C.}, \au{Demourant, F.}, \au{Poussot-Vassal, C.} \&
  \au{Sipp, D.}} \yr{2019}  \at{Linear iterative method for closed-loop control
  of quasiperiodic flows}.  \jt{J. Fluid Mech.}  \bvol{868},  \pg{26--65}.

\bibitem[Lee \& Carlberg(2020)]{Lee2020jcp}
{\sc \au{Lee, K.} \& \au{Carlberg, K.~T.}} \yr{2020}  \at{Model reduction of
  dynamical systems on nonlinear manifolds using deep convolutional
  autoencoders}.  \jt{Journal of Computational Physics}  \bvol{404},
  \pg{108973}.

\bibitem[Liu {\em et~al.\/}(2016)Liu, G\'omez \& Theofilis]{jfm:liu:2016}
{\sc \au{Liu, Q.}, \au{G\'omez, F.} \& \au{Theofilis, V.}} \yr{2016}
  \at{Linear instability analysis of low-re incompressible flow over a long
  rectangular finite-span open cavity}.  \jt{J. Fluid Mech.}  \bvol{799},
  \pg{R2 (16 pages)}.

\bibitem[Loiseau(2020)]{Loiseau2019tcfd}
{\sc \au{Loiseau, J.-C.}} \yr{2020}  \at{Data-driven modeling of the chaotic
  thermal convection in an annular thermosyphon}.  \jt{Theoretical and
  Computational Fluid Dynamics}  \bvol{34}~(4),  \pg{339--365}.

\bibitem[Loiseau \& Brunton(2018)]{jfm:loiseau:2018a}
{\sc \au{Loiseau, J.-Ch.} \& \au{Brunton, S.~L.}} \yr{2018}  \at{Constrained
  sparse {G}alerkin regression}.  \jt{J. Fluid Mech.}  \bvol{838},
  \pg{42--67}.

\bibitem[Loiseau {\em et~al.\/}(2018)Loiseau, Brunton \& Noack]{Loiseau2018}
{\sc \au{Loiseau, Jean-Christophe}, \au{Brunton, Steven~L.} \& \au{Noack,
  Bernd~R.}} \yr{2018} {\em Handbook on Model Order Reduction\/}, chap. From
  the {POD-Galerkin} method to sparse manifold models.  \publ{De Gruyter GmbH}.

\bibitem[Loiseau {\em et~al.\/}(2019)Loiseau, Bucci, Cherubini \&
  Robinet]{chapter:loiseau:2019}
{\sc \au{Loiseau, J.-Ch.}, \au{Bucci, M.~A.}, \au{Cherubini, S.} \&
  \au{Robinet, J.-Ch.}} \yr{2019}  \at{Time-stepping and {K}rylov methods for
  large-scale instability problems}.  \bt{In {\em Computational Modelling of
  Bifurcations and Instailities in Fluid Dynamics\/}},  \pg{pp. 33--73}.
  \publ{Springer}.

\bibitem[Lopez-Paz {\em et~al.\/}(2013)Lopez-Paz, Hennig \&
  Sch{\"o}lkopf]{rdc:lopez-paz:2013}
{\sc \au{Lopez-Paz, D.}, \au{Hennig, P.} \& \au{Sch{\"o}lkopf, B.}} \yr{2013}
  \at{The randomized dependence coefficient}.  \bt{In {\em Advances in Neural
  Information Processing Systems 26\/} (ed. \ed{C.~J.~C. Burges, L.~Bottou,
  M.~Welling, Z.~Ghahramani \& K.~Q. Weinberger})},  \pg{pp. 1--9}.
  \publ{Curran Associates, Inc.}

\bibitem[Lorenz(1963)]{jas:lorenz:1963}
{\sc \au{Lorenz, E.~N.}} \yr{1963}  \at{Deterministic nonperiodic flow}.
  \jt{J. Atmos. Sci.}  \bvol{20}~(2),  \pg{130--141}.

\bibitem[Lumley(1967)]{atrwp:lumley:1967}
{\sc \au{Lumley, J.~L.}} \yr{1967}  \at{The structure of inhomogeneous
  turbulent flows}.  \jt{Atmospheric turbulence and radio wave propagation} .

\bibitem[Majda \& Timofeyev(2000)]{Majda2000pnas}
{\sc \au{Majda, Andrew~J} \& \au{Timofeyev, Ilya}} \yr{2000}  \at{Remarkable
  statistical behavior for truncated {B}urgers--{H}opf dynamics}.
  \jt{Proceedings of the National Academy of Sciences}  \bvol{97}~(23),
  \pg{12413--12417}.

\bibitem[Malkus(1956)]{jfm:malkus:1956}
{\sc \au{Malkus, W. V.~R.}} \yr{1956}  \at{Outline of a theory of turbulent
  shear flow}.  \jt{J. Fluid Mech.}  \bvol{1}~(05),  \pg{521}.

\bibitem[Manti{\v{c}}-Lugo {\em et~al.\/}(2014)Manti{\v{c}}-Lugo, Arratia \&
  Gallaire]{prl:mantic:2014}
{\sc \au{Manti{\v{c}}-Lugo, V.}, \au{Arratia, V.} \& \au{Gallaire, F.}}
  \yr{2014}  \at{Self-consistent mean flow description of the nonlinear
  saturation of the vortex shedding in the cylinder wake}.  \jt{Phys. Rev.
  Lett.}  \bvol{113}~(8).

\bibitem[Maulik {\em et~al.\/}(2019)Maulik, San, Rasheed \&
  Vedula]{Maulik2019jfm}
{\sc \au{Maulik, R.}, \au{San, O.}, \au{Rasheed, A.} \& \au{Vedula, P.}}
  \yr{2019}  \at{Subgrid modelling for two-dimensional turbulence using neural
  networks}.  \jt{Journal of Fluid Mechanics}  \bvol{858},  \pg{122--144}.

\bibitem[McKeon \& Sharma(2010)]{McKeon2010jfm}
{\sc \au{McKeon, B.~J.} \& \au{Sharma, A.~S.}} \yr{2010}  \at{A critical-layer
  framework for turbulent pipe flow}.  \jt{J. Fluid Mech.}  \bvol{658},
  \pg{336--382}.

\bibitem[Meliga(2017)]{jfm:meliga:2017}
{\sc \au{Meliga, P.}} \yr{2017}  \at{Harmonics generation and the mechanics of
  saturation in flow over an open cavity: a second-order self-consistent
  description}.  \jt{J. Fluid Mech.}  \bvol{826},  \pg{503--521}.

\bibitem[Meliga {\em et~al.\/}(2009)Meliga, Chomaz \& Sipp]{Meliga2009jfm}
{\sc \au{Meliga, P.}, \au{Chomaz, J.-M.} \& \au{Sipp, D.}} \yr{2009}
  \at{Global mode interaction and pattern selection in the wake of a disk: a
  weakly nonlinear expansion}.  \jt{Journal of Fluid Mechanics}  \bvol{633},
  \pg{159--189}.

\bibitem[Mendible {\em et~al.\/}(2020)Mendible, Brunton, Aravkin, Lowrie \&
  Kutz]{Mendible2020tcfd}
{\sc \au{Mendible, A.}, \au{Brunton, S.~L.}, \au{Aravkin, A.~Y.}, \au{Lowrie,
  W.} \& \au{Kutz, J.~N.}} \yr{2020}  \at{Dimensionality reduction and
  reduced-order modeling for traveling wave physics}.  \jt{Theoretical and
  Computational Fluid Dynamics}  \bvol{34}~(4),  \pg{385--400}.

\bibitem[Mezi{\'{c}}(2013)]{arfm:mezic:2013}
{\sc \au{Mezi{\'{c}}, I.}} \yr{2013}  \at{Analysis of fluid flows via spectral
  properties of the koopman operator}.  \jt{Annu. Rev. Fluid Mech.}
  \bvol{45}~(1),  \pg{357--378}.

\bibitem[Mohebujjaman {\em et~al.\/}(2018)Mohebujjaman, Rebholz \&
  Iliescu]{Mohebujjaman2018ijnmf}
{\sc \au{Mohebujjaman, M.}, \au{Rebholz, L.~G.} \& \au{Iliescu, T.}} \yr{2018}
  \at{Physically constrained data-driven correctioon for reduced-order modeling
  of fluid flows}.  \jt{International Journal for Numerical Methods in Fluids}
  \bvol{89}~(3),  \pg{103--122}.

\bibitem[Mohebujjaman {\em et~al.\/}(2017)Mohebujjaman, Rebholz, Xie \&
  Iliescu]{Mohebujjaman2017jcp}
{\sc \au{Mohebujjaman, M.}, \au{Rebholz, L.~G.}, \au{Xie, X.} \& \au{Iliescu,
  T.}} \yr{2017}  \at{Energy balance and mass conservation in reduced order
  models of fluid flows}.  \jt{Journal of Computational Physics}
  \bvol{346}~(1),  \pg{262--277}.

\bibitem[Noack {\em et~al.\/}(2003)Noack, Afanasiev, Morzy{\'{n}}ski, Tadmor \&
  Thiele]{jfm:noack:2003}
{\sc \au{Noack, B.~R.}, \au{Afanasiev, K.}, \au{Morzy{\'{n}}ski, M},
  \au{Tadmor, G.} \& \au{Thiele, F.}} \yr{2003}  \at{A hierarchy of
  low-dimensional models for the transient and post-transient cylinder wake}.
  \jt{J. Fluid Mech.}  \bvol{497},  \pg{335--363}.

\bibitem[Noack \& Eckelmann(1994)]{jfm:noack:1994}
{\sc \au{Noack, B.~R.} \& \au{Eckelmann, H.}} \yr{1994}  \at{A global stability
  analysis of the steady and periodic cylinder wake}.  \jt{J. Fluid Mech.}
  \bvol{270}~(-1),  \pg{297}.

\bibitem[Noack {\em et~al.\/}(2011)Noack, Morzynski \& Tadmor]{book:noack:2011}
{\sc \au{Noack, B.~R.}, \au{Morzynski, M.} \& \au{Tadmor, G.}}, ed. \yr{2011}
  {\em Reduced-Order Modelling for Flow Control\/}.  \publ{Springer Vienna}.

\bibitem[Noack {\em et~al.\/}(2005)Noack, Papas \& Monkewtiz]{jfm:noack:2005}
{\sc \au{Noack, B.~R.}, \au{Papas, P.} \& \au{Monkewtiz, P.~A.}} \yr{2005}
  \at{The need for a pressure-term representation in empirical {G}alerkin
  models of incompressible shear flows}.  \jt{J. Fluid Mech.}  \bvol{523},
  \pg{339--365}.

\bibitem[Noack {\em et~al.\/}(2008)Noack, Schlegel, Ahlborn, Mutschke,
  Morzynski, Comte \& Tadmor]{Noack2008jnet}
{\sc \au{Noack, B.~R.}, \au{Schlegel, M.}, \au{Ahlborn, B.}, \au{Mutschke, G.},
  \au{Morzynski, M.}, \au{Comte, P.} \& \au{Tadmor, G.}} \yr{2008}  \at{A
  finite-time thermodynamics of unsteady fluid flows}.  \jt{Journal of
  Non-Equilibrium Thermodynamics}  \bvol{33},  \pg{103--148}.

\bibitem[\"Osth {\em et~al.\/}(2014)\"Osth, Noack, Krajnovi{\'{c}}, Barros \&
  Bor{\'{e}}e]{jfm:osth:2014}
{\sc \au{\"Osth, J.}, \au{Noack, B.~R.}, \au{Krajnovi{\'{c}}, S.}, \au{Barros,
  D.} \& \au{Bor{\'{e}}e, J.}} \yr{2014}  \at{{On the need for a nonlinear
  subscale turbulence term in {POD} models as exemplified for a
  high-Reynolds-number flow over an Ahmed body}}.  \jt{J. Fluid Mech.}
  \bvol{747},  \pg{518--544}.

\bibitem[Pan \& Duraisamy(2018)]{Pan2018siam}
{\sc \au{Pan, S.} \& \au{Duraisamy, K.}} \yr{2018}  \at{Data-driven discovery
  of closure models}.  \jt{SIAM Journal of Applied Dynamical Systems}
  \bvol{17}~(4),  \pg{2381--2413}.

\bibitem[Peherstorfer \& Willcox(2016)]{Peherstorfer2016cmame}
{\sc \au{Peherstorfer, B.} \& \au{Willcox, K.}} \yr{2016}  \at{Data-driven
  operator inference for nonintrusive projection-based model reduction}.
  \jt{Computer Methods in Applied Mechanics and Engineering}  \bvol{306},
  \pg{196--215}.

\bibitem[Picella {\em et~al.\/}(2018)Picella, Loiseau, Lusseyran, Robinet,
  Cherubini \& Pastur]{jfm:picella:2018}
{\sc \au{Picella, F.}, \au{Loiseau, J.-Ch.}, \au{Lusseyran, F.}, \au{Robinet,
  J.-Ch.}, \au{Cherubini, S.} \& \au{Pastur, L.}} \yr{2018}  \at{Successive
  bifurcations in a fully three-dimensional open cavity flow}.  \jt{J. Fluid
  Mech.}  \bvol{844},  \pg{855--877}.

\bibitem[Pickering {\em et~al.\/}(2020)Pickering, Rigas, Schmidt, Sipp \&
  Colonius]{Pickering2020}
{\sc \au{Pickering, E.}, \au{Rigas, G.}, \au{Schmidt, O.~T.}, \au{Sipp, D.} \&
  \au{Colonius, T.}} \yr{2020}  \at{Optimal eddy viscosity for resolvent-based
  models of coherent structures in turbulent jets}.  \jt{arXiv:2005.10964} .

\bibitem[Qian {\em et~al.\/}(2020)Qian, Kramer, Peherstorfer \&
  Willcox]{Qian2020}
{\sc \au{Qian, E.}, \au{Kramer, B.}, \au{Peherstorfer, B.} \& \au{Willcox, K.}}
  \yr{2020}  \at{Lift \& learn: Physics-informed machine learning for
  large-scale nonlinear dynamical systems}.  \jt{Physica D: Nonlinear
  Phenomena}  \bvol{406},  \pg{132401}.

\bibitem[Reiss {\em et~al.\/}(2018)Reiss, Schulze, Sesterhenn \&
  Mehrmann]{reiss2018shifted}
{\sc \au{Reiss, J.}, \au{Schulze, P.}, \au{Sesterhenn, J.} \& \au{Mehrmann,
  V.}} \yr{2018}  \at{The shifted proper orthogonal decomposition: A mode
  decomposition for multiple transport phenomena}.  \jt{SIAM Journal on
  Scientific Computing}  \bvol{40}~(3),  \pg{A1322--A1344}.

\bibitem[Rempfer \& Fasel(1994)]{jfm:rempfer:1994}
{\sc \au{Rempfer, D.} \& \au{Fasel, H.~F.}} \yr{1994}  \at{Dynamics of
  three-dimensional coherent structures in a flat-plate boundary layer}.
  \jt{J. Fluid Mech.}  \bvol{275}~(-1),  \pg{257}.

\bibitem[R\`enyi(1959)]{Renyi1959}
{\sc \au{R\`enyi, A.}} \yr{1959}  \at{On measures of dependence}.  \jt{Acta
  Mathematica Academiae Scientarum Hungaricae}  \bvol{10},  \pg{441--451}.

\bibitem[Rim {\em et~al.\/}(2018)Rim, Moe \& LeVeque]{Rim2018juq}
{\sc \au{Rim, D.}, \au{Moe, S.} \& \au{LeVeque, R.~J.}} \yr{2018}
  \at{Transport reversal for model reduction of hyperbolic partial differential
  equations}.  \jt{SIAM/ASA J. Uncert. Quant.}  \bvol{6}~(1),  \pg{118--150}.

\bibitem[Rossiter(1964)]{techreport:rossiter:1964}
{\sc \au{Rossiter, J.~E.}} \yr{1964}  \bt{Wind tunnel experiments on the flow
  over rectangular cavities at subsonic and transonic speeds}. {\em Tech.
  Rep.\/}.  \org{Ministry of Aviation; Royal Aircraft Establishment; RAE
  Farnborough}.

\bibitem[Rowley {\em et~al.\/}(2009{\natexlab{{\em a\/}}})Rowley, Mezi\'c,
  Bagheri, Schlatter \& Henningson]{Rowley2009}
{\sc \au{Rowley, C.}, \au{Mezi\'c, I.}, \au{Bagheri, S.}, \au{Schlatter, P.} \&
  \au{Henningson, D.~S.}} \yr{2009{\natexlab{{\em a\/}}}}  \at{Spectral
  analysis of nonlinear flows}.  \jt{Journal of Fluid Mechanics}  \bvol{641},
  \pg{115--127}.

\bibitem[Rowley {\em et~al.\/}(2002)Rowley, Colonius \& J.]{jfm:rowley:2002}
{\sc \au{Rowley, C.~W.}, \au{Colonius, T.} \& \au{J., Basu~A.}} \yr{2002}
  \at{On self-sustained oscillations in two-dimensional compressible flow over
  rectangular cavities}.  \jt{J. Fluid Mech.}  \bvol{455}.

\bibitem[Rowley {\em et~al.\/}(2004)Rowley, Colonius \& Murray]{Rowley2004pd}
{\sc \au{Rowley, Clarence~W.}, \au{Colonius, Tim} \& \au{Murray, Richard~M.}}
  \yr{2004}  \at{Model reduction for compressible flows using {POD} and
  {Galerkin} projection}.  \jt{Physica D}  \bvol{189},  \pg{115--129}.

\bibitem[Rowley \& Dawson(2017)]{Rowley2017}
{\sc \au{Rowley, C.~W.} \& \au{Dawson, S. T.~M.}} \yr{2017}  \at{Model
  reduction for flow analysis and control}.  \jt{Ann. Rev. Fluid Mech.}
  \bvol{49},  \pg{387--417}.

\bibitem[Rowley \& Marsden(2000)]{Rowley2000physd}
{\sc \au{Rowley, C.~W.} \& \au{Marsden, J.~E.}} \yr{2000}  \at{Reconstruction
  equations and the {K}arhunen--{L}o{\`e}ve expansion for systems with
  symmetry}.  \jt{Physica D: Nonlinear Phenomena}  \bvol{142}~(1-2),
  \pg{1--19}.

\bibitem[Rowley {\em et~al.\/}(2009{\natexlab{{\em b\/}}})Rowley, Mezi{\'{c}},
  Bagheri \& Schlatter]{jfm:rowley:2009}
{\sc \au{Rowley, C.~W.}, \au{Mezi{\'{c}}, I.}, \au{Bagheri, S.} \&
  \au{Schlatter, P.}} \yr{2009{\natexlab{{\em b\/}}}}  \at{Spectral analysis of
  nonlinear flows}.  \jt{J. Fluid Mech.}  \bvol{641},  \pg{115}.

\bibitem[Rubini {\em et~al.\/}(2020)Rubini, Lasagna \& Ronch]{Rubini2020jfm}
{\sc \au{Rubini, R.}, \au{Lasagna, D.} \& \au{Ronch, A.~Da}} \yr{2020}  \at{The
  {$\ell_1$}-based sparsification of energy interaction in unsteady lid-driven
  cavity flow}.  \jt{Journal of Fluid Mechanics}  \bvol{905},  \pg{A15}.

\bibitem[Ruelle \& Takens(1971)]{Ruelle1971cmp}
{\sc \au{Ruelle, D.} \& \au{Takens, F.}} \yr{1971}  \at{On the nature of
  turbulence}.  \jt{Communications in Mathematical Physics}  \bvol{20}~(3),
  \pg{167--192}.

\bibitem[San \& Maulik(2018)]{pre:san:2018}
{\sc \au{San, O.} \& \au{Maulik, R.}} \yr{2018}  \at{Extreme learning machine
  for reduced order modeling of turbulent geophysical flows}.  \jt{Phys. Rev.
  E}  \bvol{97}~(4).

\bibitem[Schlegel \& Noack(2015)]{jfm:schlegel:2015}
{\sc \au{Schlegel, M.} \& \au{Noack, B.~R.}} \yr{2015}  \at{On long-term
  boundedness of galerkin models}.  \jt{J. Fluid Mech.}  \bvol{765},
  \pg{325--352}.

\bibitem[Schmid(2007)]{Schmid2007arfm}
{\sc \au{Schmid, P.~J.}} \yr{2007}  \at{Nonmodal stability theory}.  \jt{Annual
  Review of Fluid Mechanics}  \bvol{39},  \pg{129--162}.

\bibitem[Schmid(2010)]{jfm:schmid:2010}
{\sc \au{Schmid, P.~J.}} \yr{2010}  \at{Dynamic mode decomposition of numerical
  and experimental data}.  \jt{J. Fluid Mech.}  \bvol{656},  \pg{5--28}.

\bibitem[Schmidt \& Lipson(2009)]{science:schmidt:2009}
{\sc \au{Schmidt, M.} \& \au{Lipson, H.}} \yr{2009}  \at{Distilling free-form
  natural laws from experimental data}.  \jt{Science}  \bvol{324}~(5923),
  \pg{81--85}.

\bibitem[Schmidt(2020)]{Schmidt2020bispectral}
{\sc \au{Schmidt, O.~T.}} \yr{2020}  \at{Bispectral mode decomposition of
  nonlinear flows}.  \jt{Nonlinear Dynamics}  \bvol{102},  \pg{2479--2501}.

\bibitem[Sipp \& Lebedev(2007)]{jfm:sipp:2007}
{\sc \au{Sipp, D.} \& \au{Lebedev, A.}} \yr{2007}  \at{Global stability of base
  and mean flows: a general approach and its applications to cylinder and open
  cavity flows}.  \jt{J. Fluid Mech.}  \bvol{593}.

\bibitem[Sipp {\em et~al.\/}(2010)Sipp, Marquet, Meliga \&
  Barbagallo]{amr:sipp:2010}
{\sc \au{Sipp, D.}, \au{Marquet, O.}, \au{Meliga, P.} \& \au{Barbagallo, A.}}
  \yr{2010}  \at{Dynamics and control of global instabilities in open-flows: A
  linearized approach}.  \jt{Appl. Mech. Rev.}  \bvol{63}~(3),  \pg{030801}.

\bibitem[{Sirovich}(1987)]{qam:sirovich:1987}
{\sc \au{{Sirovich}, L.}} \yr{1987}  \at{{Turbulence and the dynamics of
  coherent structures. I - Coherent structures. II - Symmetries and
  transformations. III - Dynamics and scaling}}.  \jt{Q. Appl. Math.}
  \bvol{45},  \pg{561--571}.

\bibitem[Stewart(2001)]{siam:stewart:2001}
{\sc \au{Stewart, G.~W.}} \yr{2001}  \at{A {K}rylov-{S}chur algorithm for large
  eigenproblems}.  \jt{SIAM J. Matrix Anal. Appl.}  \bvol{23},  \pg{601--614}.

\bibitem[Stuart(1958)]{Stuart1958jfm}
{\sc \au{Stuart, J.~T.}} \yr{1958}  \at{On the non-linear mechanics of
  hydrodynamic stability}.  \jt{J. Fluid Mech.}  \bvol{4}~(1),  \pg{1--21}.

\bibitem[Swinney \& Gollub(1981)]{SwinneyGollub}
{\sc \au{Swinney, H.~L.} \& \au{Gollub, J.~P.}} \yr{1981} {\em Hydrodynamic
  instabilities and the transition to turbulence\/}.  \publ{Springer-Verlag
  Berlin Heidelberg}.

\bibitem[Taira {\em et~al.\/}(2017)Taira, Brunton, Dawson, Rowley, Colonius,
  McKeon, Schmidt, Gordeyev, Theofilis \& Ukeiley]{aiaa:taira:2017}
{\sc \au{Taira, K.}, \au{Brunton, S.~L.}, \au{Dawson, S. T.~M.}, \au{Rowley,
  C.~W.}, \au{Colonius, T.}, \au{McKeon, B.~J.}, \au{Schmidt, O.~T.},
  \au{Gordeyev, S.}, \au{Theofilis, V.} \& \au{Ukeiley, L.~S.}} \yr{2017}
  \at{Modal analysis of fluid flows: An overview}.  \jt{{AIAA} J.}
  \bvol{55}~(12),  \pg{4013--4041}.

\bibitem[Tennekes \& Lumley(1972)]{Tennekes1972book}
{\sc \au{Tennekes, H.} \& \au{Lumley, J.~L.}} \yr{1972} {\em A First Course in
  Turbulence\/}.  \publ{MIT Press}.

\bibitem[Theofilis(2011)]{Theofilis2011arfm}
{\sc \au{Theofilis, V.}} \yr{2011}  \at{Global linear instability}.  \jt{Annual
  Review of Fluid Mechanics}  \bvol{43},  \pg{319--352}.

\bibitem[Towne {\em et~al.\/}(2018)Towne, Schmidt \& Colonius]{jfm:towne:2018}
{\sc \au{Towne, A.}, \au{Schmidt, O.~T.} \& \au{Colonius, T.}} \yr{2018}
  \at{Spectral proper orthogonal decomposition and its relationship to dynamic
  mode decomposition and resolvent analysis}.  \jt{J. Fluid Mech.}  \bvol{847},
   \pg{821--867}.

\bibitem[Tu {\em et~al.\/}(2014)Tu, Rowley, Luchtenburg, Brunton \&
  Kutz]{jcd:tu:2014}
{\sc \au{Tu, J.~H.}, \au{Rowley, C.~W.}, \au{Luchtenburg, D.~M.}, \au{Brunton,
  S.~L.} \& \au{Kutz, J.~N.}} \yr{2014}  \at{On dynamic mode decomposition:
  Theory and applications}.  \jt{J. Comp. Dyn.}  \bvol{1}~(2),  \pg{391--421}.

\bibitem[Vlachas {\em et~al.\/}(2018)Vlachas, Byeon, Wan, Sapsis \&
  Komoutsakos]{prsa:vlachas:2018}
{\sc \au{Vlachas, P.~R.}, \au{Byeon, W.}, \au{Wan, Z.~Y.}, \au{Sapsis, T.~P.}
  \& \au{Komoutsakos, P.}} \yr{2018}  \at{Data-driven forecasting of
  high-dimensional chaotic systems with long short-term memory networks}.
  \jt{Proc. Royal Soc. A}  \bvol{474},  \pg{20170844}.

\bibitem[Wang {\em et~al.\/}(2012)Wang, Akthar, Borggaard \&
  Ilescu]{cmame:wang:2012}
{\sc \au{Wang, Z.}, \au{Akthar, I.}, \au{Borggaard, J.} \& \au{Ilescu, T.}}
  \yr{2012}  \at{Proper orthogonal decomposition closure models for turbulent
  flows: a numerical comparison}.  \jt{Comput. Methods Appl. Mech. Eng.}
  \bvol{237-240},  \pg{10--26}.

\bibitem[Xie {\em et~al.\/}(2018)Xie, Mohebujjaman, Rebholz \&
  Iliescu]{Xie2018siam}
{\sc \au{Xie, X.}, \au{Mohebujjaman, M.}, \au{Rebholz, L.~G.} \& \au{Iliescu,
  T.}} \yr{2018}  \at{Data-driven filtered reduced order modeling of fluid
  flows}.  \jt{SIAM Journal on Scientific Computing}  \bvol{40}~(3),
  \pg{B834--b857}.

\bibitem[Yamouni {\em et~al.\/}(2013)Yamouni, Sipp \&
  Jacquin]{jfm:yamouni:2013}
{\sc \au{Yamouni, S.}, \au{Sipp, D.} \& \au{Jacquin, L.}} \yr{2013}
  \at{Interaction between feedback aeroacoustic and acoustic resonance
  mechanisms in a cavity flow: a global stability analysis}.  \jt{J. Fluid
  Mech.}  \bvol{717},  \pg{134--165}.

\bibitem[Yu(1977)]{joa:yu:1977}
{\sc \au{Yu, Y.~H.}} \yr{1977}  \at{Measurements of sound radiation from
  cavities at subsonic speeds}.  \jt{Journal of Aircraft}  \bvol{14}~(9),
  \pg{838--843}.

\end{thebibliography}
\bibliographystyle{jfm}

\end{document}